\documentclass[
12pt,
prb,
a4paper,
amsmath,
superscriptaddress,
showpacs, 
showkeys,
floatfix,
notitlepage
]{revtex4-1}

\newlength{\figwidth}

\newif\ifOnecolumn
  \ifx\Onecolumn\undefined
  \Onecolumntrue
  
\else
  \Onecolumnfalse
\fi

\newif\ifTwocolumn
  \ifOnecolumn
   \Twocolumnfalse
\else
  \Twocolumntrue
\fi

\usepackage{graphicx,amssymb,amsmath,amsfonts}
\usepackage{epsfig}
\usepackage{xcolor}
\usepackage[mathscr]{eucal}
\usepackage{comment}
\usepackage{bm}
\usepackage{xspace}

\newcommand{\latin}[1]{{\it #1}}
\newcommand{\ie}{\latin{i.e.}\@\xspace}
\newcommand{\eg}{\latin{e.g.}\@\xspace}
\newcommand{\cf}{\latin{cf.}\@\xspace}

\begin{document}

\title{Phase behaviour and structure of a superionic liquid in nonpolarized nanoconfinement}

\author{Maxym Dudka}
\email{maxdudka@icmp.lviv.ua}
\affiliation{Institute for Condensed Matter Physics, 1 Svientsitskii str., 79011 Lviv, Ukraine}

\author{Svyatoslav Kondrat}
\email{s.kondrat@fz-juelich.de}
\affiliation{Forschungszentrum J\"{u}lich, IBG-1: Biotechnology, 52425 J\"{u}lich, Germany}

\author{Alexei Kornyshev}
\email{a.kornyshev@imperial.ac.uk}
\affiliation{Department of Chemistry, Faculty of Natural Sciences, Imperial College London, SW7 2AZ, UK}

\author{Gleb Oshanin}
\email{oshanin@lptmc.jussieu.fr}
\affiliation{Sorbonne Universit\'es, UPMC Univ Paris 06, UMR 7600, LPTMC, F-75005, Paris, France}
\affiliation{CNRS, UMR 7600, Laboratoire de Physique Th\'{e}orique de la Mati\`{e}re Condens\'{e}e, F-75005, Paris, France}

\begin{abstract}

The ion-ion interactions become exponentially screened for ions confined in ultranarrow metallic pores. To study the phase behaviour of an assembly of such ions, called  a \emph{superionic} liquid, we develop a statistical theory formulated on bipartite lattices, which allows an analytical solution within the Bethe-lattice approach. Our solution predicts the existence of ordered and disordered phases in which ions form a crystal-like structure and a homogeneous mixture, respectively. The transition between these two phases can potentially be first or second order, depending on the ion diameter, degree of confinement and pore ionophobicity. We supplement our analytical results by three-dimensional off-lattice Monte Carlo simulations of an ionic liquid in slit nanopores. The simulations predict formation of ionic clusters and ordered snake-like patterns, leading to characteristic close-standing peaks in the cation-cation and anion-anion radial distribution functions.

\end{abstract}


\date{\today}

\maketitle

\section{Introduction}

The rejuvenation of interest to fundamental mechanisms of energy storage in electric double-layer capacitors (also called supercapacitors) has been boosted by the development of novel materials for nanostructured electrodes\cite{frackowiak:07, miller:sci:08, simon:nm:08, simon_gogotsi:acr:13, yoo:nanolett:11:graphene, yang:sci13:graphene, lukatskaya:sci:13, naguib:am:13:2D, gogotsi:nm:15:2D} and by a booming research in room temperature ionic liquids.\cite{earle:pac:00, galinski:ea:06, endres:02, buzzeo:cpc:06, ohno:book} This progress in material science has been accompanied by detailed investigations of performances of such systems. In particular, pioneering experimental studies~\cite{gogotsi:sci:06, pinero:carbon:06, chmiola:angewandte:08, gogotsi:08} have demonstrated that using  electrodes with ultranarrow pores, able to accommodate about one layer (or row) of ions, leads to a substantial increase of the \emph{surface-specific} capacitance. This `anomalous' increase of capacitance for subnanometer pores can be explained by a \emph{superionic state} emerging in such metal-like pores:~The ion-ion interactions become exponentially screened, and this allows an easier packing of ions of the same type. An improved mean-field model has been developed that shows that the superionic state leads ultimately to higher capacitances for narrower pores.~\cite{kondrat:jpcm:11} Many aspects of this theory have later been verified by computer simulations.\cite{kondrat:pccp:11, feng:pccp:11, merlet:natmat:12, wu:qiao:jpcl:12, xing:jpcl:trans:13} 

Subsequent works have focused on voltage-dependent capacitances,~\cite{kondrat:pccp:11, wu:qiao:jpcl:12, kondrat:ees:12} optimization of energy storage~\cite{kondrat:ees:12, xing_bedrov:jpcl:12, vatamanu:jpcl:energystorage:13, merlet:natcom:13} and dynamics of charging.\cite{kondrat:jpcc:13, lee:nanotech:14, kondrat:nm:14, pean:acsnano:14, he:jpcl:15} Surprisingly, however, the structure and phase behaviour of an ionic liquid in nanoconfinement have received much less attention so far, and we know only about a voltage-induced phase transition between dilute and dense phases predicted by theory\cite{kondrat:jpcm:11, lee:16:hyster} or seen in simulations.\cite{ kiyohara:jcp:11, xing:jpcl:trans:13,vatamanu:acsnano:15} On the other hand, for flat electrodes there is experimental evidence~\cite{lockett:jpcc:hyster:08, zhou:ec:hysterflat:10, druechler:jpcc:hyster:10, uysala:jpcc:hyster:13} of hysteretic behaviour of capacitance, whose origin is not yet clear, while simulations~\cite{merlet:jpcc:14:EDLOwnLifeTransition, rotenberg_salanne:jpcl:15} suggest  a structural transition between \emph{ordered} and disordered states in the interfacial region of an ionic liquid at flat electrodes. We shall demonstrate in this work that the ordered state should also exist in nano-confinement, show a possibility of a phase transition to a homogeneous mixture of ions (preferable for fast charging), and elaborate on the structure of ionic liquids in narrow slit pores. We restrict our considerations to \emph{non-polarized} pores, setting the basis for the study of voltage-dependent behaviour, which we defer to future works, however.

\begin{figure}
\begin{center}
\includegraphics[width=.4\hsize]{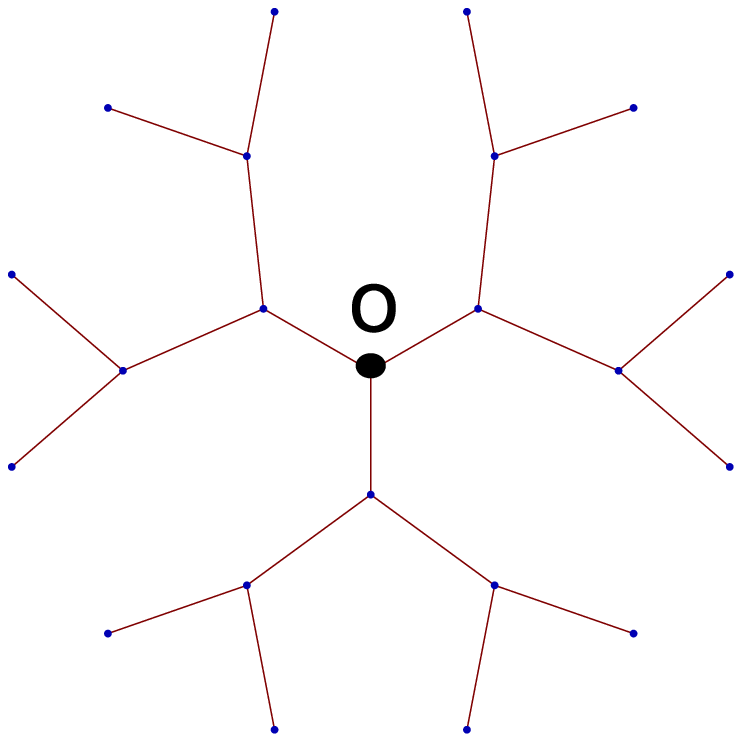}
\includegraphics[width=.4\hsize]{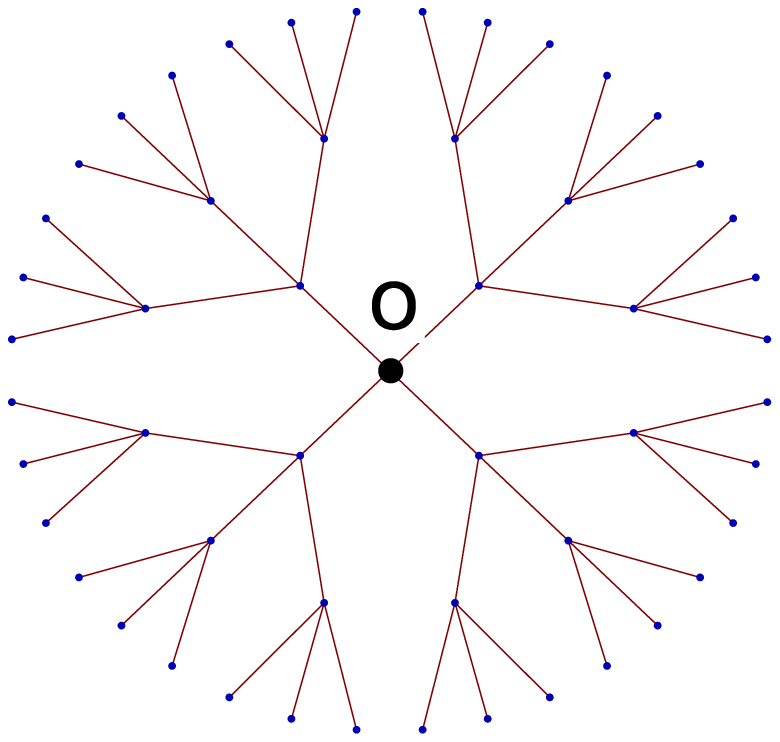}

\hspace{0.5cm}\mbox{(a)}\hspace{6cm}\mbox{(b)}
\caption{Schematic of the Cayley tree with coordination numbers $q=3$ (a) and $q=4$ (b) and $N=3$ generations emanating from the central (or root) site denoted by `O'. The bond length between the nodes of the Caylee tree is the same and appears different for aesthetic reasons only. We shall use the Cayley tree to obtain an analytical solution for a lattice model of an ionic liquid in slit nanopores.
}
\label{Fig0}
\end{center}
\end{figure}

Analytically tractable models are among the most precious assets in physics, as they often allow to trace system properties exactly such that new physical insights can be more easily developed.~\cite{Baxter} In the context of ionic liquids and supercapacitors, examples include a lattice model for dense ionic liquids;~\cite{kornyshev:07} a one-dimensional Coulomb lattice capacitor;~\cite{Horgan1DIsingCaoacitor:2012, Horgan1DIsingOverscreening:2012, demery_podgornik:2015:1dCLG} one-dimensional Ising,~\cite{kornyshev:fd:14} Blume-Emery-Griffiths~\cite{lee:prl:14, rochester:1d} and harmonic oscillator~\cite{schmickler:ea:2015:harmonicOscillator} models adapted for single-file pores; and the already mentioned continuous mean-field model for ultrathin slit pores.~\cite{kondrat:jpcm:11, lee:16:hyster} 

In this work we introduce a lattice model for a \emph{superionic liquid}, \ie an ionic liquid in the superionic state, in slit nanoconfinement. Our model can be directly mapped onto the standard Blume-Emery-Griffiths or Blume-Capel models (Appendix~\ref{app:map2BEG}), and is solved here using the Bethe-lattice approach for bipartite lattices with coordination numbers (numbers of the nearest neighbours) $q=3$ and $q=4$ (Section~\ref{sec:bethe} and Appendix~\ref{sec:q4}, respectively). By definition, the Bethe lattice represents a deep interior of the so-called Cayley tree (a structure consisting of $q$ branches emanating from a central or root site and having $N$ generations, see Figure~\ref{Fig0}), discarding the effects of the boundary sites and thus describing system's bulk properties. It is important to note that as any other tree graph the Bethe lattice can be partitioned into two sublattices only.~\cite{scheinerman:book} For lattices that can partition into a larger number of sublattices (\eg triangular lattice, which is tripartite), the Bethe lattice approach is not applicable and one has to resort to other approximations.\cite{zhuk} 

Although the Bethe lattice approach is unlikely to predict correct critical exponents, it has proven to give qualitatively and quantitatively accurate predictions for the location and order of phase transitions. Examples of this are numerous and include athermal lattice gases,~\cite{runnels,brazil} modulated phases of the Ising model with competing interactions,~\cite{van,hor,mariz} Potts models,~\cite{potts} lattice models of glassy systems~\cite{biroli1,biroli2} and localisation transitions.~\cite{biroli3} Thus, there is a good reason to believe that the Bethe-lattice approach used here will describe correctly the phase behaviour of ions in slit nanoconfinement.

We supplement our analytical results by three dimensional off-lattice Monte Carlo simulations of an ionic liquid in ultranarrow slit pores.~\cite{kondrat:pccp:11} Our simulations unravel further details in the system behaviour and reveal the structure that the ionic liquid adopts in such a strong nanoconfinement (Section~\ref{sec:mc}). 

Finally, we will conclude and critically discuss our results in Section~\ref{sec:concl}.

\section{Lattice model of a superionic liquid}
\label{sec:bethe}

We consider an ionic liquid (IL) confined in a slit metallic nanopore so narrow that only one IL layer can fit in it. We assume that the ions reside on the symmetry plane of this pore and consider a 2D lattice occupied by cations (+), anions (-) or voids. The occupation of site $i$ can be described by a pair of Boolean variables
\begin{equation}
(n_i,m_i) = \left\{\begin{array}{ll}
(1,0),     \mbox{  site $i$ is occupied by  `$+$' particle,} \nonumber\\
(0,1),      \mbox{  site $i$ is occupied by `$-$' particle,} \\
(0,0),      \mbox{  site $i$ is empty}, \nonumber
\end{array}
\right.
\end{equation}
where the case $(1,1)$ is excluded because the ions cannot occupy the same site due to hard core interactions.

The partition function of this system in thermal equilibrium is given by
\begin{equation}
Z=\prod_i \sum_{(n_i,m_i)}\exp[-\beta \mathcal{H}],
\end{equation}
where $\beta=1/k_B T$ is the reciprocal temperature measured in units of the Boltzmann constant $k_B$, and the Hamiltonian is given by
\begin{eqnarray}\label{hamil}
{\mathcal H}=\sum_{\langle ij\rangle} \Big[ I_{++} n_i n_j + I_{--} m_i m_j-I_{+ -}(n_i m_j+n_j m_i) \Big]
	- \sum_{i} \Big[\mu_+^{(0)} n_i+\mu_-^{(0)} m_i \Big].
\end{eqnarray}
The summation in (\ref{hamil}) extends over all pairs of \emph{nearest-neighbouring} sites $\langle ij\rangle$ and $I_{\alpha\beta}>0$ denote the strengths of the respective interactions, where $\alpha, \beta=\{+,-\}$. Since the inter-ionic Coulomb potential is exponentially screened in narrow metallic pores,~\cite{kondrat:jpcm:11} we have taken into account only nearest-neighbour interactions in this work, but we note that it would be interesting to study the effects due to the next-to-nearest (or higher) neighbour interactions too. The electrochemical potentials of cations and anions (in infinite dilution) are 
\begin{equation}
\label{eq:bethe:mu}
\mu_{\pm}^{(0)}=\pm e V+w_{\pm},
\end{equation}
where $V$ is the electrostatic potential of the pore walls measured with respect to the IL bulk and $w_{\pm}$ are energies of transfer of $\pm$ ions from the pore interior into the bulk. In general case $w_+\neq w_-$. We note that $w_\pm$ include the ion-pore wall interactions due to the image forces (see Section~\ref{sec:sim:method} and Eq.~(\ref{eq:self})), and that they are here defined such that a large positive $w_\pm$ corresponds to a cationo/aniono-philic pore.~\footnotemark[1]

Throughout the paper we assume that the coupling constants, $I_{\alpha\beta}$, do not depend on the occupation numbers $(n_i, m_i)$; we shall make a similar assumption in our Monte Carlo simulations as well, where we assume that the pore width does not change as the ions enter the pore (see Section~\ref{sec:mc}). It is clear however that intrusion of sufficiently large ions into a narrow pore can create stresses on the pore and potentially expand it.~\cite{hantel:ec:11:dilatometry, kaasik_presser:ec:13:ElectrodeSwelling, rochester_kornyshev:ea:15:UnwantedActuation, hantel:14:carbon} This may particularly occur as a response to the applied voltage (not studied in the present work, however). Such `unwanted electroactuation' is detrimental to supercapacitors, which should ideally function without any mechanical stress,\cite{rochester_kornyshev:ea:15:UnwantedActuation} and deserves further studies. However, incorporation of such stresses into the present model would lead to serious complications, and the problem would become analytically untreatable. We shall therefore neglect such effects in our present work, and note that swelling/contraction of pores is typically of the order of a few percents,\cite{rochester_kornyshev:ea:15:UnwantedActuation} and it seems reasonable to expect that it will have only a minor effect on our results.

The model defined by Eq.~(\ref{hamil}) can be mapped onto the Blume-Emery-Griffiths model, which has been intensively studied by a  large variety of methods. We discuss this mapping in Appendix~\ref{app:map2BEG}. In the next subsection, we solve our model for $V=0$ and $w_+=w_-$ using the Bethe-lattice approximation. The readers not interested in the details of our analytical approach can switch directly to subsection~\ref{sec:bethe:res}, where we discuss the results and physical implications of our solution.

\subsection{Bethe-lattice solution}
\label{sec:bethe:sol}

Let us  consider a Cayley tree (see Figure~\ref{Fig0}) with the coordination number $q$; that is, a central, root site with $q$ branches emanating from it and having each $N$ generations of sites. The partition function of our model of an ionic liquid on such a Cayley tree can be represented as
\begin{equation}\label{pfd}
Z_N=Z_N(0)+Z_N(+)+Z_N(-),
\end{equation}
where $Z_N(0)$ is an auxiliary partition function of a Cayley tree with a vacant central site, while
 $Z_N(+)$ and $Z_N(-)$ are auxiliary constrained partition functions  of  the model with the root site occupied by a particle '+' or occupied by a particle  '--', respectively. In what follows, 
we will derive general recursion relations obeyed by these auxiliary functions and then turn to the Bethe-lattice solutions of these recursions.

The Cayley tree can be cut apart at the central site into $q$ identical branches. Therefore, we have
\begin{equation}\label{zzz}
Z_N(0)=g^q_N(0),\quad Z_N(+)=z_{+}g^q_N(+),\quad Z_N(-)=z_{-}g^q_N(-),
\end{equation}
where $z_{\pm}=\exp \beta w_{\pm}$ and $g_N(0)$ and $g_N(\pm)$ are the partition functions of one branch with initial vacant cite and the branches with initial sites occupied by particle `$+$' or `$-$' respectively. Each branch consists of $q-1$ identical sub branches. Therefore, we can  write the following recursion relations for $g_N$'s:
\begin{eqnarray}\label{recc}
g_N(0)&=&g^{q-1}_{N-1}(0)+z_{+}g^{q-1}_{N-1}(+)+z_{-}g^{q-1}_{N-1}(-),\nonumber\\
g_N(+)&=&g^{q-1}_N(0)+z_{+}e^{-\beta I_+}g^{q-1}_{N-1}(+)+z_{-}e^{\beta I_{+-}}g^{q-1}_{N-1}(-),\nonumber\\
g_N(-)&=&g^{q-1}_N(0)+z_{+}e^{\beta I_{+-}}g^{q-1}_{N-1}(+)+z_{-}e^{-\beta I_-}g^{q-1}_{N-1}(-).
\end{eqnarray}
Next, introducing new variables :
\begin{equation}\label{nvar}
x_N=\frac{g_N(+)}{g_N(0)},
\quad 	\textrm{and} \quad 
y_N=\frac{g_N(-)}{g_N(0)}
\end{equation}
we obtain a system of two coupled  recursion relations :
\begin{eqnarray}\label{recq}
x_N&=&\frac{1+z_{+}e^{-\beta I_+}x^{q-1}_{N-1} +z_{-}e^{\beta I_{+-}}y^{q-1}_{N-1}}{1+z_{+}x^{q-1}_{N-1}+z_{-}y^{q-1}_{N-1}}\nonumber\\
y_N&=&\frac{1+z_{+}e^{\beta I_{+-}}x^{q-1}_{N-1} +z_{-}e^{-\beta I_-}y^{q-1}_{N-1}}{1+z_{+}x^{q-1}_{N-1}+z_{-}y^{q-1}_{N-1}}
\end{eqnarray}
Mean densities of particles `$+$' or `$-$'  on the central site of the Cayley tree can then be straightforwardly expressed via the variables $x_N$ and $y_N$ as
\begin{equation}\label{dense}
\rho_{0,+}=\frac{z_{+}x^{q}_N}{1+z_{+}x^{q}_N+z_{-}y^{q}_N},\quad
\rho_{0,-}=\frac{z_{-}y^{q}_N}{1+z_{+}x^{q}_N+z_{-}y^{q}_N} \,.
\end{equation}

We turn next to the behavior in the interior part
of the  Cayley tree and the limit $N\to\infty$, \ie, on the Bethe lattice, for which all sites are equivalent and hence,  all  $\{x_N,y_N\}$ should  converge to a fixed point or cycle solutions  $\{x,y\}$. We start with the calculation of the free energy, which contains some subtleties, since the effect of the boundary sites has to be correctly  excluded  (see, e.g., discussion in Ref.~\onlinecite{Gujrati95}). To this end, we use here the following procedure 
elaborated for generalized recursive lattices \cite{Ananikian98}.
Substituting (\ref{zzz}) into (\ref{pfd}) and taking into account (\ref{nvar})  we have
\begin{equation}\label{pfz}
Z_N=g_N^q(0)\left(1+z_{+}x^{q}_N+z_{-}y^{q}_N\right) \,,
\end{equation}
so that the free energy of $N$-generation Cayley tree can be cast into the form
\begin{equation}
-\beta F_N=\ln Z_N=q\ln g_N(0) +\ln\left(1+z_{+}x^{q}_N+z_{-}y^{q}_N\right) \,.
\end{equation}
Next, for $g_N(0)$ we use first relation of (\ref{recc}), which gives, together  with the definitions in (\ref{nvar}) :
\begin{equation}
-\beta F_N=q(q-1)\ln g_{N-1}(0)+q\ln\left(1+z_{+}x^{q-1}_{N-1}+z_{-}y^{q-1}_{N-1}\right) +\ln\left(1+z_{+}x^{q}_N+z_{-}y^{q}_N\right)
\end{equation}
Rewriting the latter expression as
\begin{eqnarray}
-\beta F_N &=& -(q-1) \beta F_{N-1}
	-(q-1)\ln\left(1+z_{+}x^{q}_{N-1}+z_{-}y^{q}_{N-1}\right) + \nonumber\\
 	&& q\ln\left(1+z_{+}x^{q-1}_{N-1}+z_{-}y^{q-1}_{N-1}\right) +\ln\left(1+z_{+}x^{q}_N+z_{-}y^{q}_N\right),
\end{eqnarray}
and repeating this procedure $n$ times,  we  arrive at the following recursion relation obeyed by the free energy :
\begin{equation}
-\beta F_N=-(q-1)^n\beta F_{N-n}-\beta F_{Nn}
\end{equation}
where the last term is the free energy of the $n$-generation Cayley tree 
\begin{multline}
-\beta F_{Nn} = q \sum_{k=1}^n(q-1)^{k-1} \ln\left(1+z_{+}x^{q-1}_{N-k}+z_{-}y^{q-1}_{N-k}\right) \\
	- (q-1)^n \ln\left(1+z_{+}x^{q}_{N-n}+z_{-}y^{q}_{N-n}\right)
	+ \ln\left(1+z_{+}x^{q}_N+z_{-}y^{q}_N\right)
\end{multline}
Further on, in the limit $N\to \infty$ all $x_{N-k} \equiv x$ and $y_{N-k} \equiv y$, so that
\begin{eqnarray}
-\beta F_{n}=\lim_{N\to\infty}(-\beta F_{Nn})&=& q\frac{(q-1)^n-1}{q-2}\ln\left(1+z_{+}x^{q-1}+z_{-}y^{q-1}\right)-\nonumber\\&&
((g-1)^n-1)\ln\left(1+z_{+}x^{q}+z_{-}y^{q}\right)
\end{eqnarray}
To obtain the free energy per site, one should divide the latter expression by the 
number of bulk sites $N_{s}$, comprising the Bethe lattice, in $n$-generation Cayley tree. According to Gujrati \cite{Gujrati95},  $N_s$ is simply related  to the number of bonds $N_b$ via the homogeneity assumption $N_b/N_s=q/2$, and in $n$-generation Cayley tree we have
\begin{equation}
N_b=q\frac{(q-1)^n-1}{q-2} \,,
\end{equation}
and hence,
\begin{equation}\label{free}
-\beta f=-\frac{\beta F_{n}}{N_s}= \frac{q}{2}\ln\left(1+z_{+}x^{q-1}+z_{-}y^{q-1}\right)-
 \frac{q-2}{2}\ln\left(1+z_{+}x^{q}+z_{-}y^{q}\right).
\end{equation}

We note that similar calculations for the Ising model lead to the free energy which is equivalent to the free energy obtained by the integration of the equation of state.\cite{Baxter}

We focus now  on the Bethe lattices with the coordination numbers $q=3$ and $q=4$ in the completely symmetric case when $I_+=I_-=I_{+-}=I$ and $w_+=w_-$ (that is $z_+=z_-$) which is appropriate to the model of ionic liquids in non-polarised confinement (more precisely, for potential of zero charge) and for ions of the same size and with the same interaction with the pore walls. In this symmetric case, the original Blume-Emery-Griffiths model reduces to a simpler Blume-Capel model in a magnetic field (see Appendix~\ref{app:map2BEG} for more details).
We present below a detailed derivation of the results for the lattice with the coordination number $q = 3$, while the analogous derivation for the case $q =4$, which shows the same qualitative behavior, is discussed in Appendix~\ref{sec:q4}.

\subsubsection{Solution for the Bethe lattice with coordination number $q=3$}
\label{sec:bethe:q3}

For $q=3$ our recursion relations in (\ref{recq}) take the  form
\begin{eqnarray}
\label{rec}
x_N&=&\varphi(x_{N-1},y_{N-1}) \nonumber\\
y_N&=&\varphi(y_{N-1},x_{N-1})
\end{eqnarray}
with
\begin{equation}\label{function}
\varphi({x,y})=\frac{1+z (e^{-\beta I}x^{2} +e^{\beta I} y^{2})}{1+z(x^{2}+y^{2})} \,.
\end{equation}
We  note that, generally speaking, the recursion scheme presented above has been already studied in the past. However, all the previous analysis was focused solely  on the case of the ferromagnetic Blume-Capel (BC) model. It is not clear \emph{a priori} if the results of this analysis will still hold for our case ($I > 0$) which corresponds to the antiferromagnetic BC model. Hence, we find it expedient to derive the explicit solution here.

\begin{figure}
\includegraphics[width=.3\hsize]{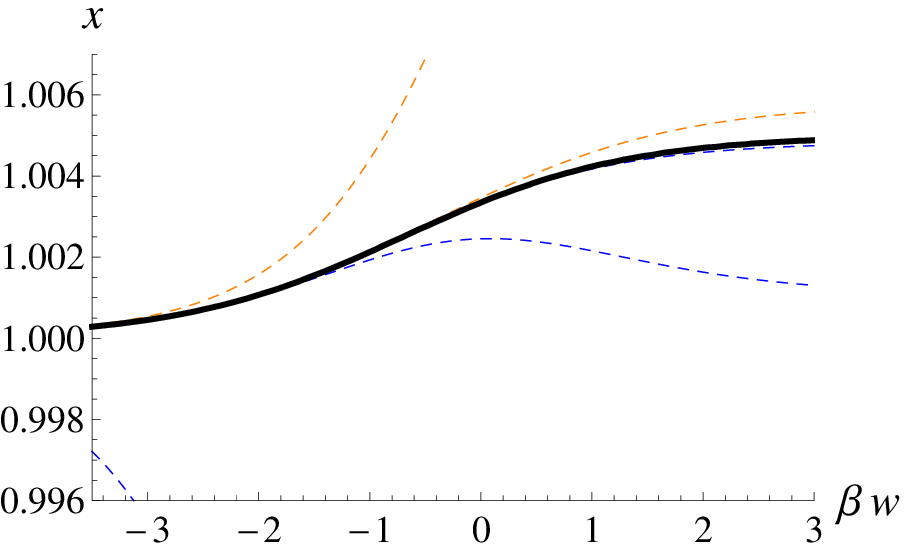} 
\includegraphics[width=.3\hsize]{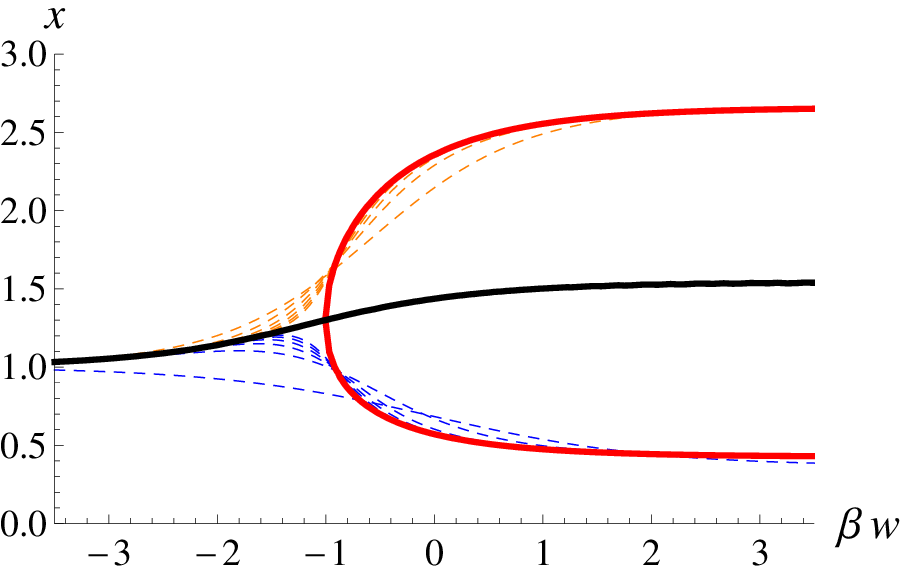} 
\includegraphics[width=.3\hsize]{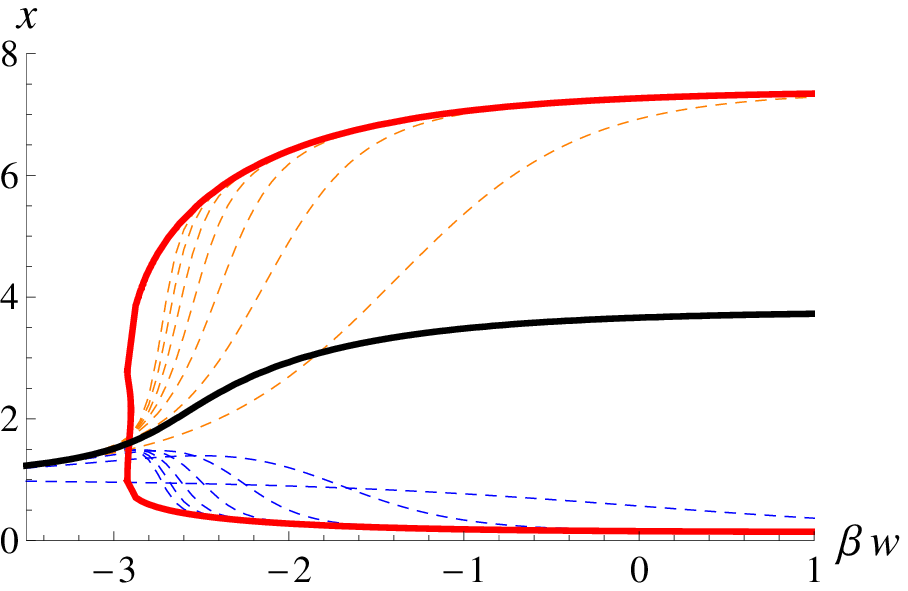}
\includegraphics[width=.3\hsize]{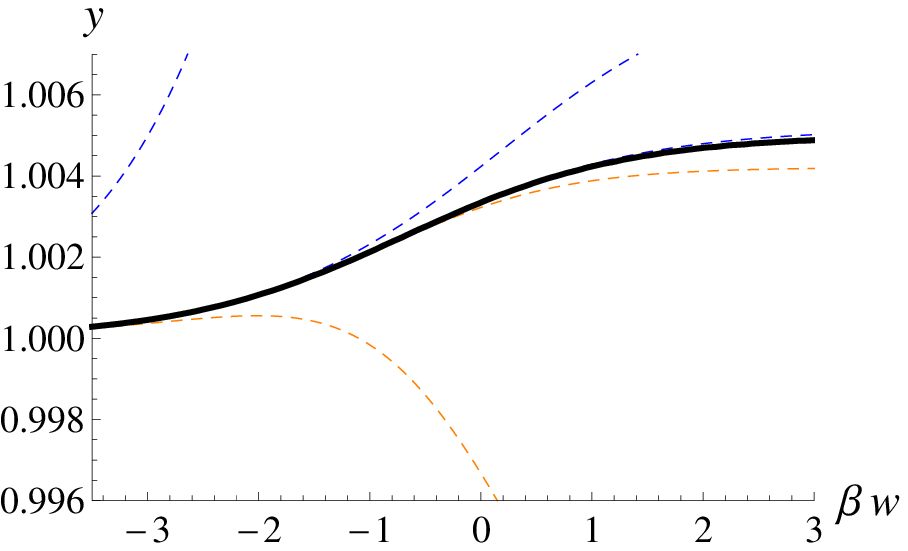} 
\includegraphics[width=.3\hsize]{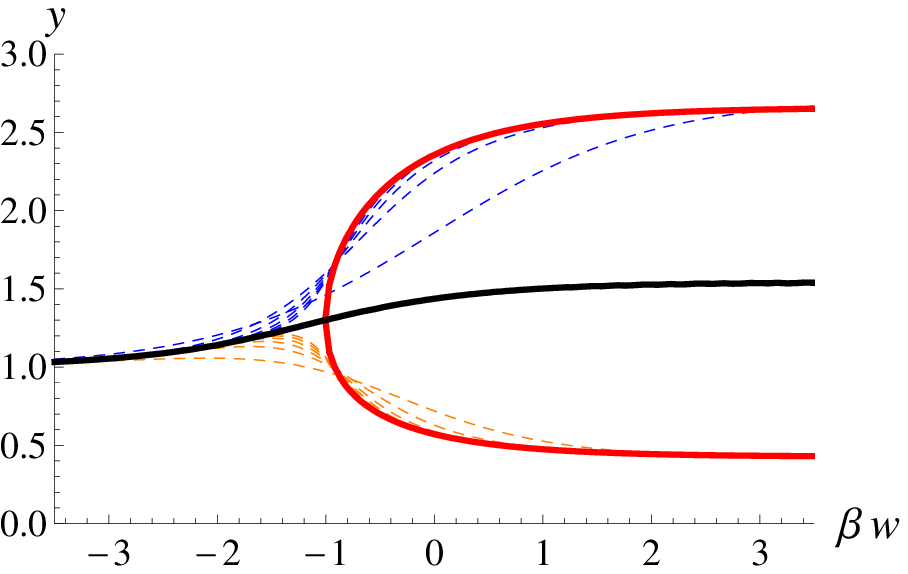} 
\includegraphics[width=.3\hsize]{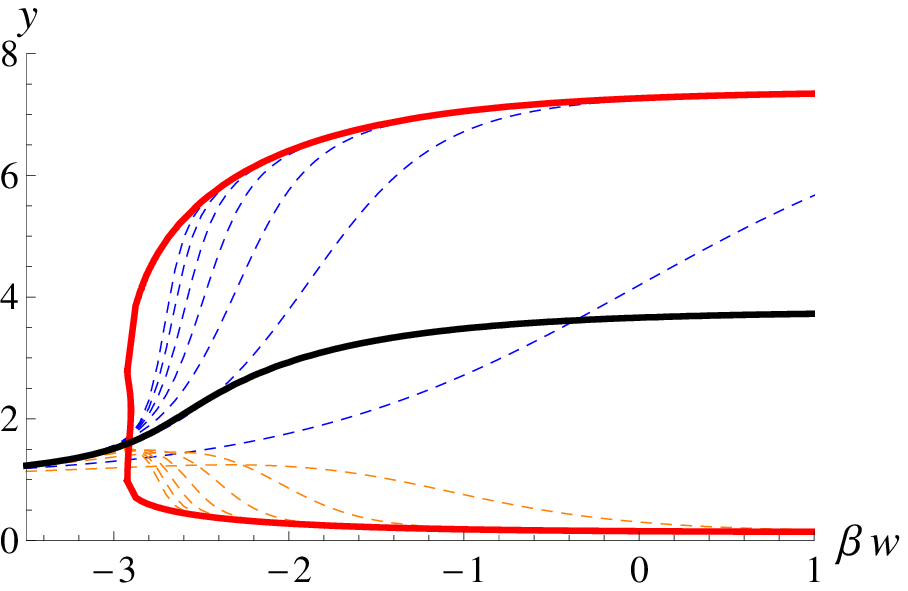}

\hspace{0.5cm}\mbox{(a)}\hspace{5cm}\mbox{(b)}\hspace{5cm}\mbox{(c)}
\caption{Spontaneous breaking of symmetry between sub-lattices. Thick lines show the limiting solutions ($N \to \infty$) of the recursion relations (\ref{rec}), $x$ and $y$. Thin dash lines denote the first few terms in the recursion, $x_N$ and $y_N$, for $N=1,2,\cdots$, which were obtained by starting from $x_0 = 1$ and $y_0=0$ (not shown) in all cases; these lines approach the corresponding solid lines as $N$ increases. The top row shows $x$ and $x_N$, and the bottom row $y$ and $y_N$, as functions of the resolvation energy $\beta w$ for (a) $\beta I=0.1$, (b) $\beta I=1$ and (c) $\beta I=2$. (a)~$(x_N, y_N)$  converge to a single solution $(x,y)$ for any $\beta w$. (b)~A single solution exists only for $\beta w < (\beta w)_\mathrm{tr} \approx -1$. It becomes unstable above $(\beta w)_\mathrm{tr}$, where there are additionally two solutions for which $x_{2N}$ (dash orange lines) and $x_{2N+1}$ (dash blue lines) converge to different values as $N\to \infty$, and similarly $y_{2N}$ and $y_{2N+1}$. These limiting solutions are shown by thick red lines. They describe spontaneous breaking of symmetry between two sub-lattices, so that the sub-lattices with odd and even $N$ have different ion densities in equilibrium, determined by these two solutions (note that $x_N$ and $y_N$ are related to the ion densities by Eq.~(\ref{denseA})). Physically it means that for $\beta w > (\beta w)_\mathrm{tr}$ the system is in the ordered, symmetry broken state, in which the ion densities on different sub-lattices are different, reminiscent of a crystalline structure (\cf Figure~\ref{fig:honeycomb}). Below $(\beta w)_{\mathrm{tr}}$ and in case (a), the system is in a homogeneous state characterized by the same average ion density on all sites. The transition between these two states is second order, as discussed in the text (\cf also Figure~\ref{fig:f1f4}a). (c)~The same as (b) but the transition is discontinuous (first order). This is because each solution is multivalued close to $(\beta w)_\mathrm{tr} \approx -3$, manifesting metastable states (see the upper and lower branches of the red curves close to $(\beta w)_\mathrm{tr}$, \cf also Figure~\ref{fig:f1f4}b). 
}
\label{Fig1}
\end{figure}

\begin{figure}
\includegraphics[width=.3\hsize]{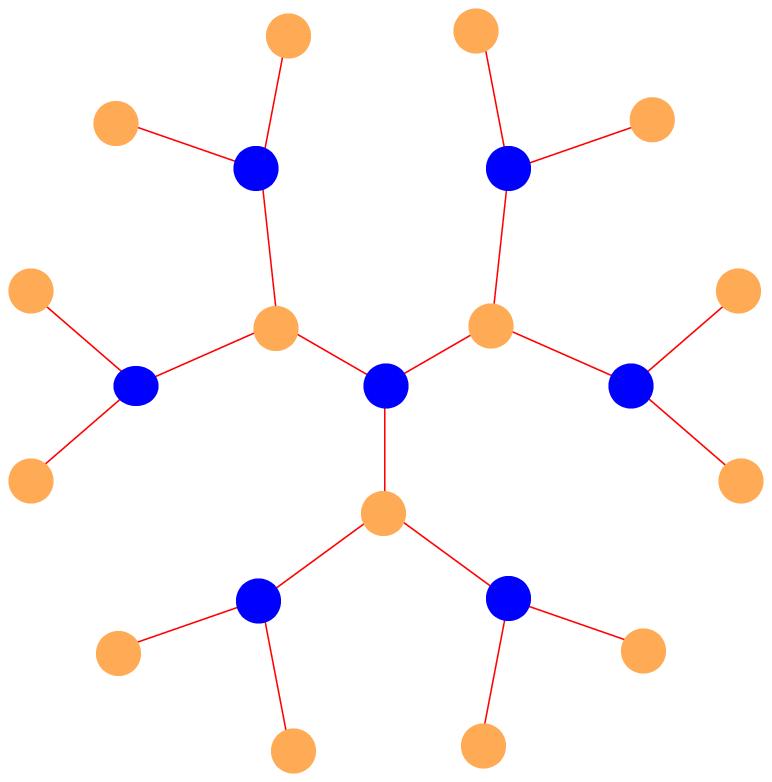}
\includegraphics[width=.27\hsize]{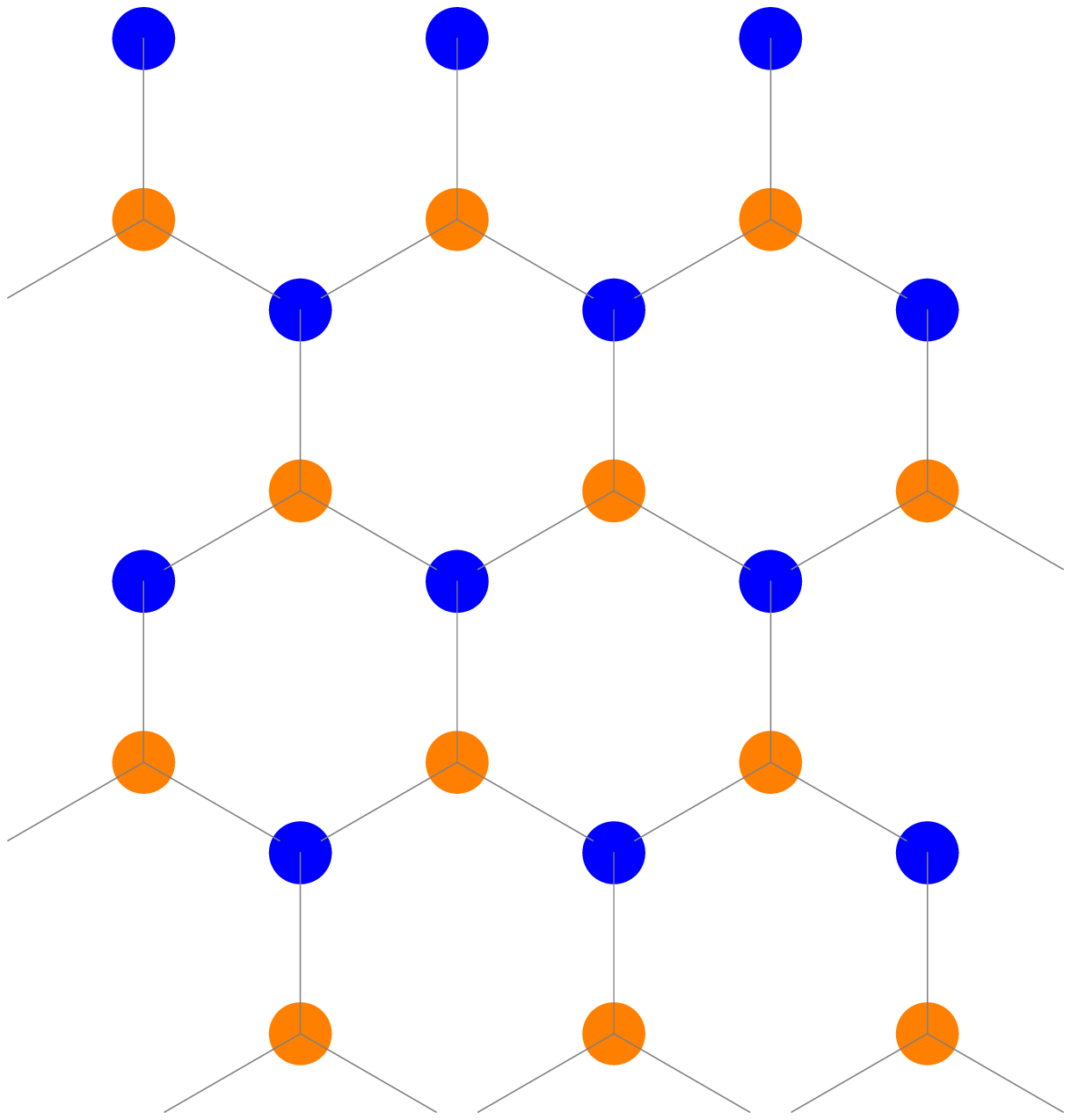}

\hspace{0.5cm}\mbox{(a)}\hspace{5cm}\mbox{(b)}
\caption{Schematic of completely ordered states corresponding to the infinite resolvation energies, $\beta w \to \infty$. (a) A fragment of the Bethe lattice with the coordination number $q = 3$ where cations (orange) and anions (blue) occupy alternating layers of odd and even generations, $N$, of the Cayley tree, respectively. (b) The corresponding fragment of the honeycomb lattice with $q = 3$. Cations and anions occupy sub-lattices shifted by one lattice spacing with respect to each other. The properties of a system defined on a honeycomb lattice are described well by the Bethe-lattice approach with $q=3$ (see \textit{e.g.}~Ref.~\onlinecite{morita:pla:83}). Our Bethe-lattice solution reveals the existence of an ordered state, the extreme case of which is shown in panel (a).}
\label{fig:honeycomb}
\end{figure}

To get a hint on the behavior of $x_N$ and $y_N$, we first generate  several consecutive  terms for $x_N$ and $y_N$ by merely iterating (\ref{rec}). These terms, as functions of $\beta w$, are depicted in Figure~\ref{Fig1}. As one may readily observe, for small $\beta I$ there is only one symmetrical solution for which all $x_N$ and $y_N$ converge to some $N$-independent curves $x$ and $y$.  However, as $\beta I$ exceeds some critical value, one observes an apparent symmetry breaking  so that $x_N$ (and $y_N$) with $N$ odd and even converge to different $N$-independent functions. This means  that for sufficiently large $\beta I$ the recursion scheme in (\ref{rec}) has cycle solutions with period $2$. This is a direct consequence of the bipartite nature of the Bethe lattice. Physically, it means that for such $\beta I$ the system looses its homogeneity and spontaneously partitions into two subsystems with the behavior of the observables on these sub lattices being different from each other.  For homogeneous regular lattices, it means that the systems partitions into two sub lattices shifted with respect to each other by one lattice spacing. For the Bethe lattice it means that it partitions into sublattices composed of layers with with even and odd number $N$ (see Figures~\ref{Fig0} and \ref{fig:honeycomb}). 

To take into account this partitioning into two sublattices,  we recall the classical analysis by Runnels \cite{runnels} of the phase diagram of a single-species mixture of identical hard molecules on the Bethe lattice, and rewrite relations (\ref{rec}) in the thermodynamical limit $N\to\infty$ in the form:
\begin{align}
x_A =\varphi(x_{B},y_{B}) \nonumber\\
y_A= \varphi(y_{B},x_{B})  \label{A}
\end{align}
and
\begin{align}
x_B =\varphi(x_{A},y_{A}) \nonumber\\
y_B =\varphi(y_{A},x_{A}) \label{B}
\end{align}
where $x_A$ and $y_A$ ($x_B$ and $y_B$) denote variables describing sublattice A (B), respectively. Once we stipulate that our central site is in sublattice A, we can write densities of '+'s and '--'s defined in (\ref{dense})
as 
\begin{equation}
\label{denseA}
\rho^A_{+}=\frac{z x^{3}_B}{1+z(x^{3}_B+y^{3}_B)},\quad
\rho^A_{-}=\frac{z y^{3}_B}{1+z(x^{3}_B+y^{3}_B)}
\end{equation}
In turn, if we stipulate that the central site  belongs to the sublattice B, the expressions for the densities in this case can be obtained from (\ref{denseA}) by a mere interchange of sub- and superscripts A and B.
 
To describe ordering on sub-lattices, we introduce 
\begin{equation}
\label{eq:op:def}
\bar\rho^A=(\rho^A_{+}+\rho^A_{-}), \qquad
\delta\rho^A=(\rho^A_{+}-\rho^A_{-}).
\end{equation}
Evidently, the difference $ \delta\rho^A$ between the  density of particles '+' and the density  of particles '--'  is the order parameter, while $\bar\rho^A$ is the total density of all particles on the sub-lattice $A$.

Substituting equation (\ref{B}) into (\ref{A}) we have, formally,
\begin{align}
x_A = \varphi(\varphi(x_{A},y_{A}),\varphi(y_{A},x_{A})) \nonumber\\
y_A = \varphi(\varphi(y_{A},x_{A}),\varphi(x_{A},y_{A}))  \label{full}
\end{align}
Note that these equations can be  also obtained from  (\ref{rec})  by iterating these equations twice
to involve the numbers of generations $N$ having the same parity (see, e.g., the discussion in Ref.~\onlinecite{runnels}) and than taking the limits $x_N\to x_A$ and $y_N\to y_A$ . 
 
\begin{figure}
\includegraphics[width=.42\hsize]{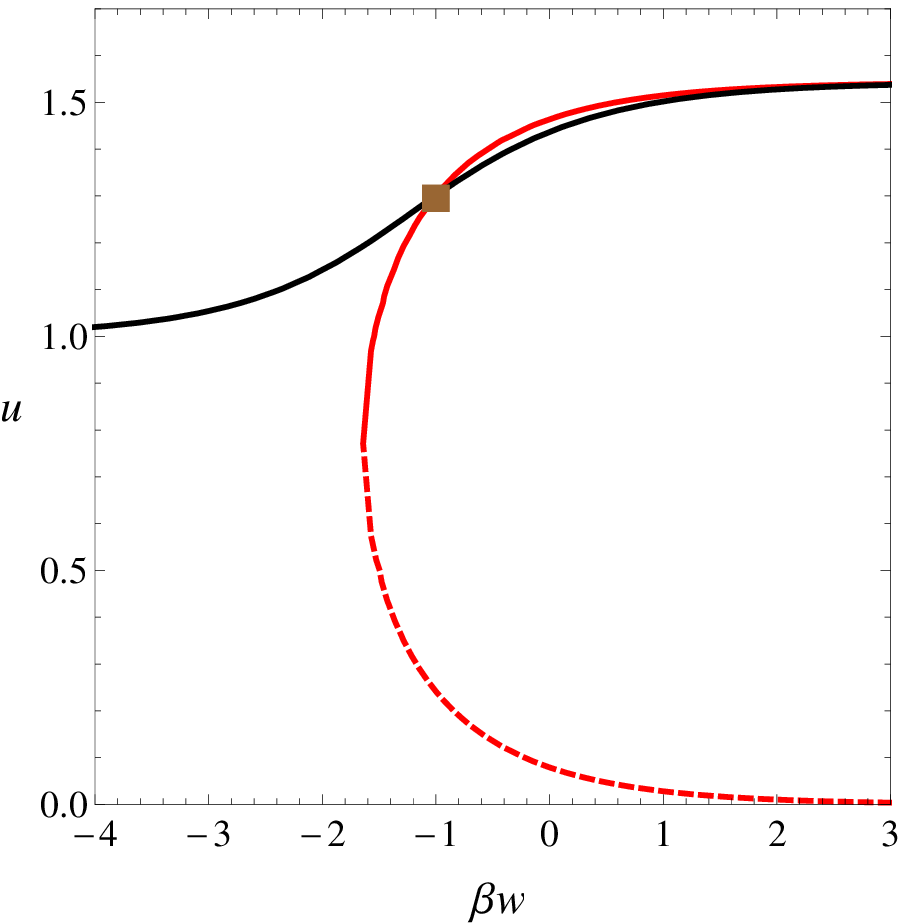} 
\includegraphics[width=.4\hsize]{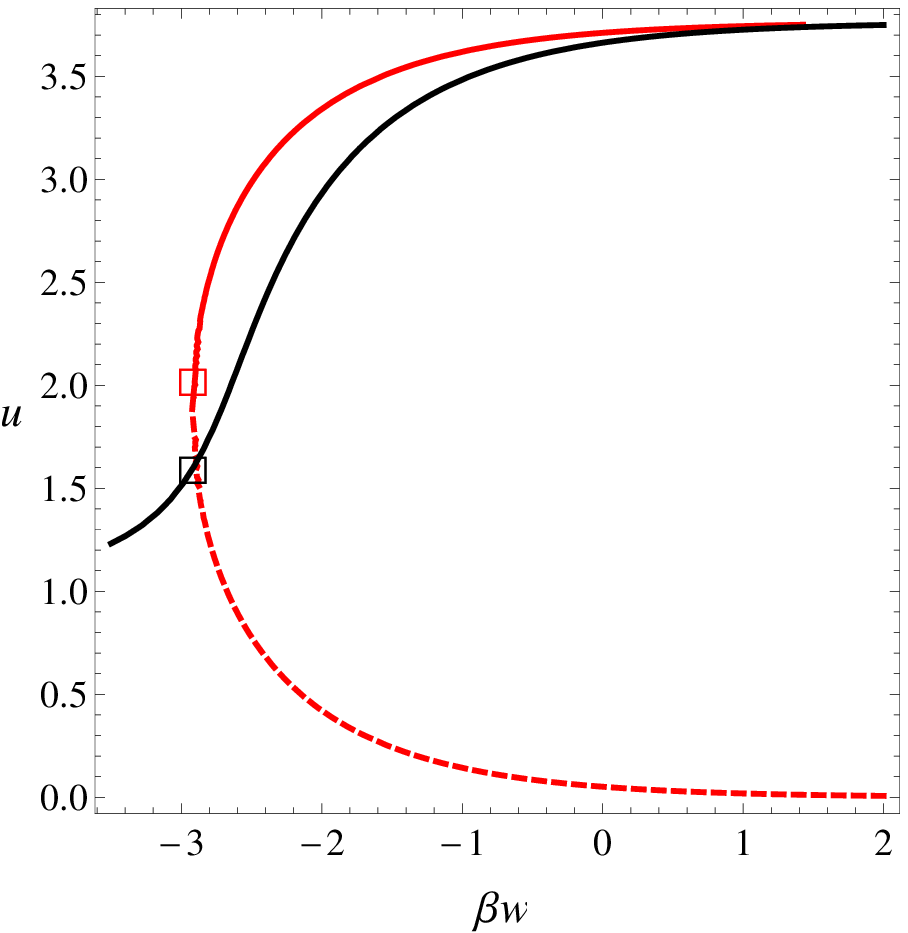}

\hspace{1cm}\mbox{(a)}\hspace{6cm}\mbox{(b)}
\caption{Locating phase transition points. Solution of $f_1(u,\beta w) = 0$ (Eq.~\eqref{eq:f1}, black line) and $f_4(u,\beta w) = 0$ (Eq.~\eqref{eq:f4}, red line) for  (a) $\beta I=1$ and (b) $\beta I=2$. Functions $f_1$ and $f_4$ describe extrema of the free energy. The lower branches of the $f_4=0$ curve (dash red lines) correspond to a decrease of $u = (x_A + y_A)/2$ with increasing $\beta w$ and are unstable. Panel (a) corresponds to a second order phase transition (see also Figure~\ref{Fig1}a). Here the $f_1 = 0$ solution describing the disordered phase (black line) intersects  with the upper branch of the $f_4 = 0$ solution describing the ordered phase (red line), giving a critical point at $(\beta w)_c=-1.0046$ for $\beta I=1$ (filled square in panel (a)). With increasing  $\beta I$ the intersection point slides down on the upper branch of the $f_4 = 0$ solution (black line) passing an extremum $\delta w / \delta u = 0$ corresponding to a tricritical point; this happens at $(\beta I)_\mathrm{tc} \approx 1.8$ and $(\beta w)_\mathrm{tc} \approx -2.61$ (not shown in this plot, but see Eqs.~(\ref{eq:3crit}) and Figure~\ref{fig:phase_diagram}). Panel (b) shows the case $\beta I = 2 > (\beta I)_\mathrm{tc}$ and hence corresponds to a first order phase transition (see also Figure~\ref{Fig1}c). The transition is determined by the equality of the free energies calculated along the solutions $f_1 = 0 $ and $f_4 = 0$. These points are denoted by open squares in panel (b). For the whole phase diagram see Figure~\ref{fig:phase_diagram}.
\label{fig:f1f4}
}
\end{figure}

It is convenient to introduce
\begin{align}
\label{eq:u_v}
u=\frac{x_A+y_A}{2}
	\quad\textrm{and}\quad 
v=\frac{x_A-y_A}{2}.
\end{align}
These variables are related to the total ion density, $\bar \rho$, and the order parameter, $\delta \rho$, as given by Eqs.~(\ref{eq:op:def}) and (\ref{eq:uv}). The order parameter describes the excess of one sort of ions on one of the two sublattices, and  hence $v=0$, implying $\delta \rho = 0$ (see Eq.~(\ref{eq:uv})), corresponds to the homogeneous (or disordered) state of ions.

Now, using Eqs.~(\ref{full}) and (\ref{function}), it can be shown that there are two stable thermodynamic phases (there are in total four functions extremizing the free energy, but only two of them lead to physically correct solutions, see Appendix~\ref{sec:q3:details} for details):
\begin{enumerate}
\item \emph{Disordered phase} described by $v=0$, with $u$ satisfying
\begin{align}
\label{eq:f1}
f_1(u,z,I) \equiv u - 1 + 2 u^3 z \left (1 - \cosh(\beta I)\right) = 0.
\end{align}

\item \emph{Ordered phase} described by $v\ne 0$, with $u$ and $v$ satisfying
\begin{subequations}
\label{eq:f4}
\begin{align}
f_4(u,z,I) &\equiv 1 - \cosh(\beta I) - 4 u z \sinh(\beta I) \left( u-\cosh(\beta I) \right) = 0, \\
4u z \sinh(\beta I) &= 1+2z\left(u^2+v^2\right).
\end{align}
\end{subequations}

\end{enumerate}
Thus, equations (\ref{eq:f1}) and (\ref{eq:f4}) describe extrema of the free energy and hence determine the transitions between these two phases. This is illustrated in Figure~\ref{fig:f1f4} for $\beta I=1$ and $\beta I=2$. For $\beta I=1$ (Figure~\ref{fig:f1f4}a), the intersection of the solution of $f_1(u,z,I) = 0$  with the upper branch of the solution of $f_4(u,z,I) = 0$ corresponds to a critical point~\cite{Ananikian91} (see also Figure~\ref{Fig1}a). A line of critical points can thus be calculated from the system of equations
\begin{eqnarray}
\label{eq:f1f4eq0}
f_1(u,z,I)&=&0 ,\nonumber\\
f_4(u,z,I)&=&0 .
\end{eqnarray}
This leads to an implicit equation for the critical temperature $(\beta I)_{c}$
\begin{equation}
z_c = \exp (\beta w)_c = \frac{ (-2 \sinh ({(\beta I)_c})+\cosh ({(\beta I)_c})-1)^2}{8\sinh^2((\beta I)_c) (2 \sinh ({(\beta I)_c})-\cosh ({(\beta I)_c}))}
\end{equation}
The solution of this equation is shown by a solid line in Figure~\ref{fig:phase_diagram}. Note that as $z\to\infty$ (strongly ionophilic pores), $\beta I$ approaches a constant value $(\beta I)_\mathrm{threshold}=\ln (\sqrt{3})$, which implies that a transition into the ordered phase can only take place for $\beta I$ above this threshold value. We also note parenthetically that it can be calculated for arbitrary coordination number $q$ of the Bethe lattice and is given by $(\beta I)_\mathrm{threshold}=\ln (\sqrt{q/(q-2)})$, see Refs.~\onlinecite{Chakraborty86,Ananikian91}. Consequently, $(\beta I)_\mathrm{threshold}$ is a monotonically \emph{ decreasing} function of $q$, meaning that for bipartite lattices with a larger coordination number, the transition into the ordered phase will occur at lower values of the resolvation energy $w$, at a given $\beta I$, and at lower values of $\beta I$, for a given $z = \exp(\beta w)$.

Visually comparing the behavior of the recursions $x_N$ and $y_N$ for two different values of $\beta I$, presented in Figure~\ref{Fig1}, one may notice that  for  $\beta I=2$ the recursion (\ref{rec}) converges to the limiting solutions $x$ and $y$ more abruptly than it happens for $\beta I=1$. Moreover, one sees that in the former case the  solutions depicted by the red line become multivalued, which signals that the transitions to the ordered phase may have a different order. Indeed, for the former case we have a first order transition with a discontinuous behavior of the density, while in the latter case the transition is continuous, with a jump in the compressibility.  This implies in turn that the line of critical points terminates at a tricritical point  $((\beta|I|)_{tc},z_{tc})$. According to Ref.~\onlinecite{Ananikian91} this tricritical point is given by 
\begin{align}
\left.\frac{\delta z}{\delta u} \right|_{v=0} =
	 \left.\frac{\partial z}{\partial u} \right|_{v=0} +
	 \left.\frac{\partial z}{\partial v^2}\frac{\partial v^2}{\partial u} \right|_{v=0} = 0
\end{align}
Formally it corresponds to the condition that the solutions of equations $f_1(u,z,I) = 0$ and $f_4(u,z,I) =0$ intersect each other exactly at the extremum of $f_4(u,z,I)$, see Figure~\ref{fig:f1f4}; this means
\begin{align}
\label{eq:3crit}
f_1(u,z,I) = 0,\nonumber\\
f_4(u,z,I) = 0, \\
\frac{\partial f_4(u,z,I)}{\partial u} = 0. \nonumber
\end{align}
The solution of Eqs.~(\ref{eq:3crit}) is $(\beta|I|)_{tc} \approx 1.8$ and $z_{tc} \approx 0.07$ (corresponding to $\beta w_{tc} = \ln z_{tc} \approx -2.61$), and is shown by a filled circle in Figure~\ref{fig:phase_diagram}. The value of $u$ at the tricritical point is $u \approx 1.56$ giving $\bar \rho = 0.359 $ (see Eqs.~(\ref{eq:uv})).

The line of the first order phase transitions for $z<z_{tc}$ (or equivalently for $\beta w < \beta w_{tc}$) can be found in the usual fashion by matching the free energies calculated for disordered and ordered phases. This is shown in Figure~\ref{fig:f1f4}b where the values of $\beta w$ and $u$ at a transition are depicted by open squares. 

Finally, it is interesting to note that for lattices that can partition into a larger number of sub-lattices (for instance a triangular lattice, which is tripartite), the behavior of the Blume-Capel model (see Appendix~\ref{app:map2BEG}) is more  intricate. In particular, in the ordered state the system partitions into three sub-lattices, two of which are ordered and predominantly occupied by ions of one type, with the third sub-lattice remaining in the `disordered'  state (\ie the order parameter is zero, see Ref.~\onlinecite{zhuk} and references therein). As mentioned in the introduction, this behaviour cannot be captured by the Bethe-lattice approach and will not be discussed in the present work.

\subsection{Bethe-lattice results}
\label{sec:bethe:res}

\begin{figure}
\includegraphics[width=.7\hsize]{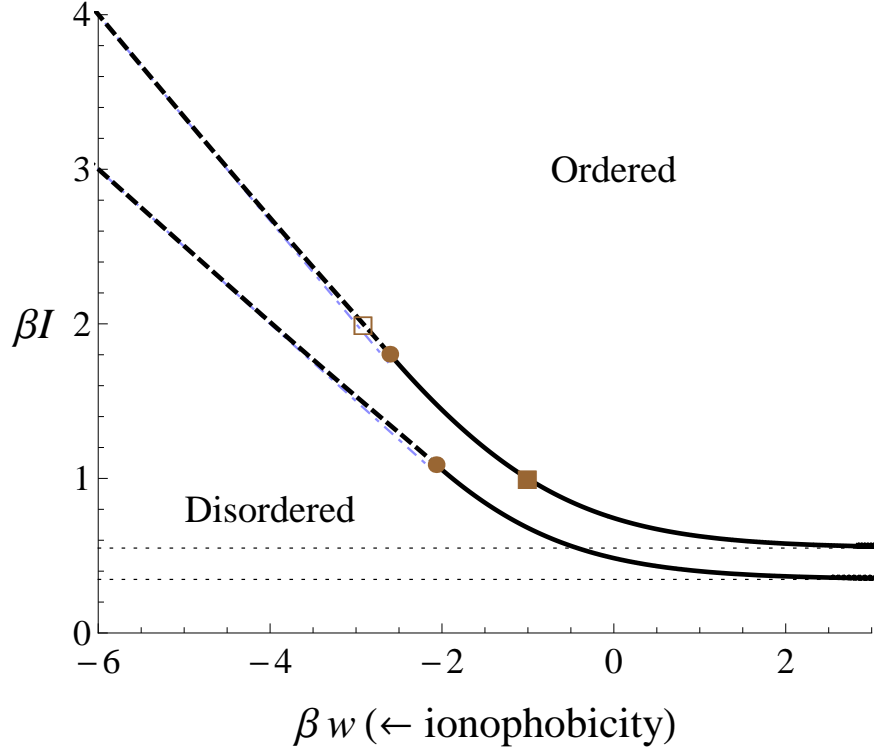}

\caption{Phase diagram of a superionic liquid in a non-polarised nanoconfinement obtained by the Bethe-lattice approximation. The diagram is plotted in the ($\beta I, \beta w$) plane, where $I$ is the pore-width dependent interaction strength (\cf Figure~\ref{fig:pot}) and $w = w_\pm$ the ion's resolvation energy, defining the ionophobicity of pores and determining their occupation by ions at zero voltage. The disordered phase is a homogeneous mixture of ions of two types and voids, and the ordered phase means that the ions of one type predominantly occupy one of the `sub lattices' (\ie ions form a crystal-like structure, see Figure~\ref{fig:honeycomb}, \cf Figure~\ref{fig:mcsim:struct}d). The upper (lower) lines correspond to coordination number $q=3$ ($q=4$), saying how many nearest neighbours has an ion. The solid lines show second order and dash lines first order phase transitions. These lines meet at tricritical points ($(\beta w)_{tc} \approx -2.61$ and $(\beta|I|)_{tc} \approx 1.8$ for $q=3$, and $(\beta w)_{tc} \approx -2.07$ and $(\beta|I|)_{tc} \approx 1.09$ for $q=4$) depicted by filled circles. Thin dash horizontal lines indicates the value of $\beta I$ below which the ordered phase does not exist ($(\beta I) = 1/2\ln ({3})$ for $q=3$ and $(\beta I) = 1/2\ln ({2})$ for $q=4$, see text). Filled and open squares denote the values of $\beta I$ and $\beta w$ considered in Figures~\ref{fig:f1f4}a and b, respectively. Thin blue lines show the lines $I = -2 w / q$ which describe the first order transitions in the limit $w\to \infty$ (see Section~\ref{sec:concl}). 
}
\label{fig:phase_diagram}
\end{figure}

We first briefly summarize the previous subsection. Our analytical solution reveals two stable thermodynamic phases: An ordered phase, where cations and anions mainly reside on different sub lattices, forming a crystal-like structure; and a homogeneous or disordered phase in which ions and voids form a homogeneous mixture. We were able to determine the location of a phase transition between these two phases and to identify its order for coordination numbers (numbers of the nearest neighbours) $q=3$ and $q=4$. This is summarized in Figure \ref{fig:phase_diagram} in the form of a phase diagram in the ($\beta w, \beta I$) plane, where $w=w_\pm$ is ion's resolvation energy (defined here as an energy of transfer of an ion from the pore into the bulk of a supercapacitor\footnotemark[1]); and $I$ is the pore-width dependent strength of the screened ion-ion interactions. In this figure, the dash line corresponds to a first order and the solid line denotes a second order phase transition. These two types of transitions meet at a tricritical point, $((\beta w)_{tc}, (\beta I)_{tc})$, denoted by filled circles in Figure~\ref{fig:phase_diagram}.

\begin{figure}
\includegraphics[width=.42\hsize]{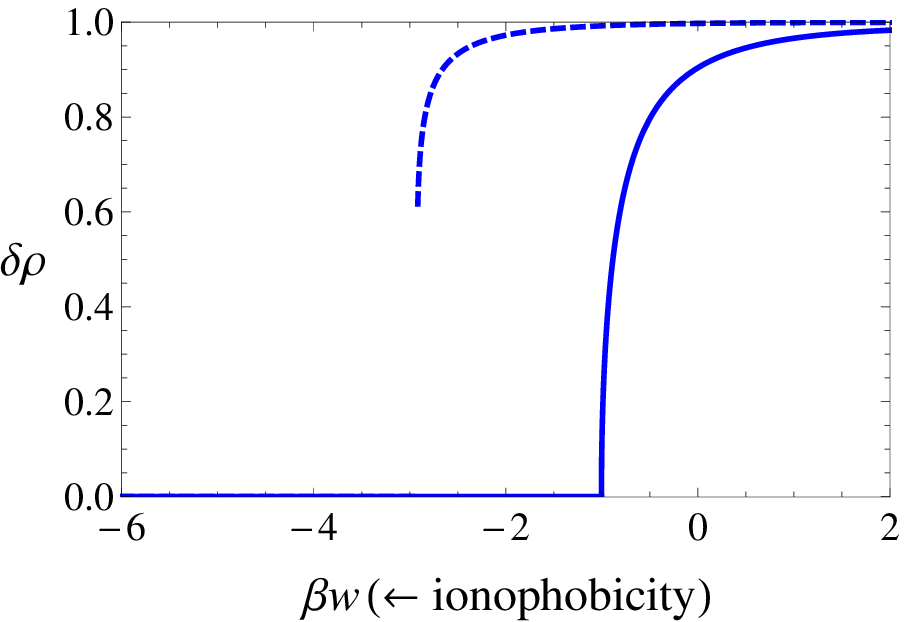} 
\includegraphics[width=.42\hsize]{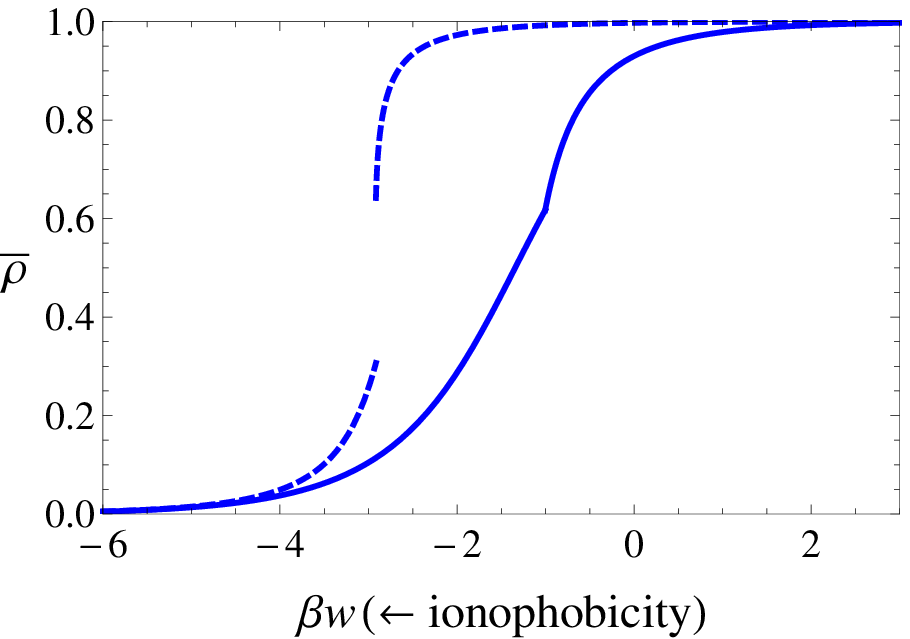}

\hspace{1cm}\mbox{(a)}\hspace{6cm}\mbox{(b)}
\caption{
(a)~Order parameter, $\delta\rho$, and (b)~total ion density, $\bar\rho$, for the coordination number $q=3$ as functions of resolvation energy $\beta w$ obtained within the Bethe-lattice approach. The solid lines are calculated for the ion-ion interaction strength $\beta I=1$, at which we observe a second order (continuous) phase transition between the ordered and disordered states. In this case both quantities are continuous but exhibit a cusp at the transition. The dash lines are for $\beta I=2$ at which the transition is first order. Here $\delta \rho$ and $\rho$ exhibit a finite jump at the transition. For the complete phase diagram see Figure~\ref{fig:phase_diagram}. 
}
\label{fig:bethe:op}
\end{figure}

\begin{figure}
\includegraphics[width=.42\hsize]{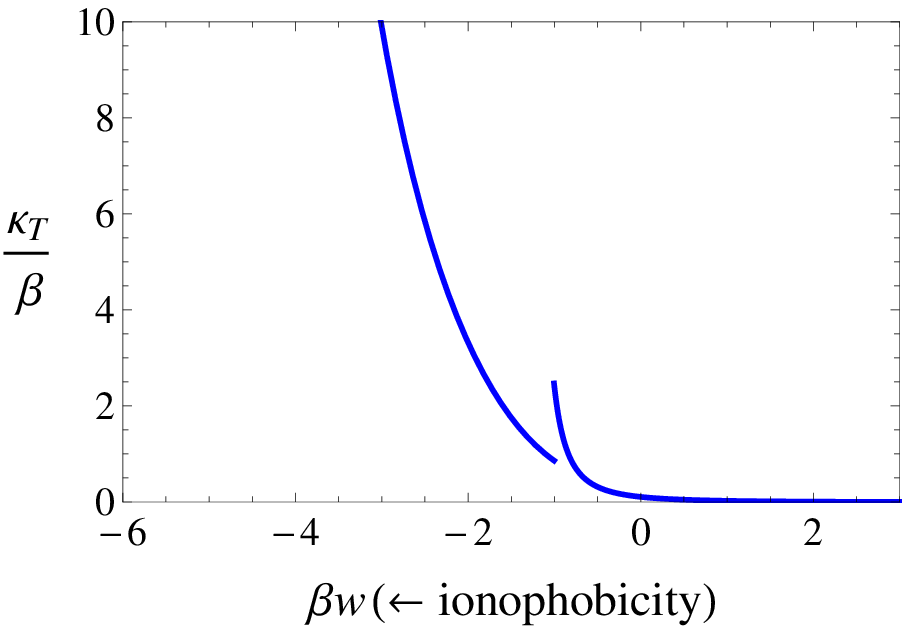} 
\includegraphics[width=.42\hsize]{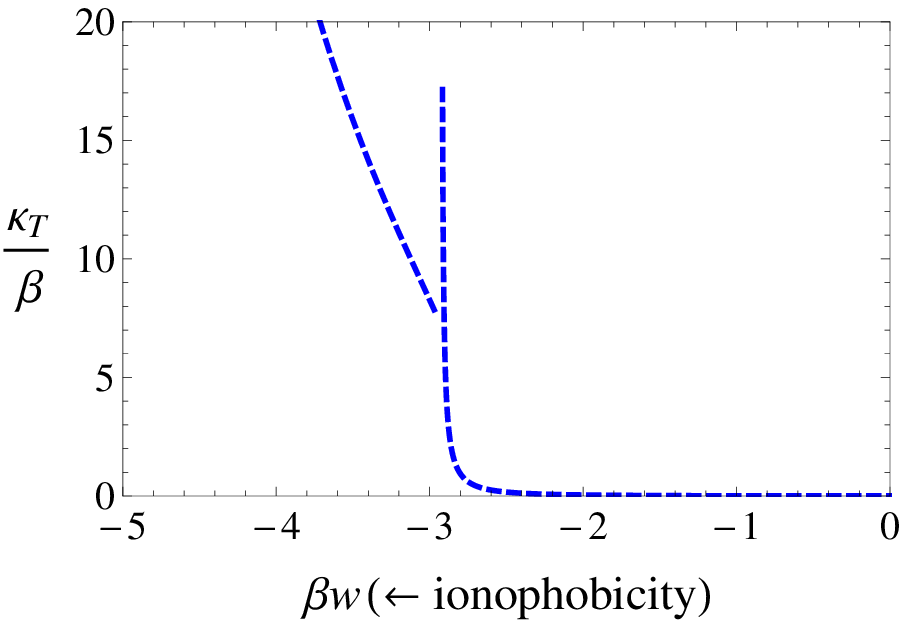}

\hspace{1cm}\mbox{(a)}\hspace{6cm}\mbox{(b)}
\caption{Analog of compressibility defined as $\kappa_T = (1/\bar \rho^2 ) \partial \bar \rho / \partial w$ is plotted as a function of resolvation energy $\beta w$ obtained within the Bethe-lattice approach. (a)~In the case of second order phase transitions, $\kappa_T$ experiences a finite jump at the transition. (b) For first order transitions, $\kappa_T$ behaves discontinuously as well, but there is a spike at the transition due to the jump in the total ion density (the dash line in Figure~\ref{fig:bethe:op}b), which is however not shown here for clarity. The values of the strength of the ion-ion interaction are $\beta I=1$ in (a) and $\beta I=2$ in (b). For the plots of the total ion density ($\bar \rho$) see Figure~\ref{fig:bethe:op}b, and Figure~\ref{fig:phase_diagram} for the complete phase diagram.
}
\label{fig:compres}
\end{figure}

The orders of these phase transitions are directly related to the behaviours of the order parameter, $\delta \rho$, and total ion density, $\bar \rho$, at the transition (see Eq.~(\ref{eq:op:def}) for definition and note that we skip the sublattice index $A$ due to the $A \leftrightarrow B$ symmetry). Both quantities vanish for strongly ionophobic pores (large negative $\beta w$) and increase to unity when the ionophilicity increases (large positive $\beta w$), \ie~the ion density increases and the system becomes more ordered for increasing $\beta w$. As usual, however, for a second order phase transition, the order parameter is a continuous function of $\beta w$ with a cusp at a transition, while it exhibits a finite jump in the case of the first order transitions observed for $\beta I$ above the tricritical point $(\beta I)_{tc}$ (solid and dash lines in Figure~\ref{fig:bethe:op}a, respectively). The behaviour of the total ion density ($\bar \rho$) is similar and is shown in Figure~\ref{fig:bethe:op}b for completeness.

Interestingly, Figure~\ref{fig:bethe:op} (dash lines) shows a very steep decrease of the density and order parameter at the first order transition. This suggest a small metastability window and hints that the transition may be only weakly first order. Further research is needed to resolve this issue, however.

An important signature of any phase transition is an analog of compressibility, defined here as $\kappa_T=(1/\bar \rho^2)\partial \bar\rho \big/\partial w$. Although it is not clear whether this quantity can be directly extracted from experiments, we present its analysis because it provides additional information about our phase transitions. For the second order transitions, occurring for $(\beta I)_\mathrm{threshold} < (\beta I) < (\beta I)_{tc}$, $\kappa_T$ shows a finite jump at the transition (Figure~\ref{fig:compres}a). Above the trictirical point, the transition is a first order and there is a spike in $\kappa_T$ at the transition (not shown) due to the jump in the total ion density (the dash line in Figure~\ref{fig:bethe:op}b), while the discontinuity in $\kappa_T$ becomes very large (Figure~\ref{fig:compres}b). 

Taking a few $k_BT$ as a typical value of $\beta I$ for room-temperature ionic liquids (\cf Figure~\ref{fig:pot}) and assuming conventional \emph{ionophilic} pores, corresponding to positive $w$, we deduce from Figure~\ref{fig:phase_diagram} that our superionic liquid must be in the ordered state under normal conditions. This is consistent with recent molecular dynamics simulations showing a crystal-like structure of ions in narrow slit pores at no applied voltage.\cite{kondrat:nm:14} We shall look at the structure of an ionic liquid in slit nanopores in a slightly more detail in the next section, where we discuss the results of our three-dimensional off-lattice Monte Carlo simulations.

\section{Off-lattice Monte Carlo simulations}
\label{sec:mc}

In addition to the analytical results based on the Bethe-lattice approach, we present the results of grand canonical off-lattice Monte Carlo simulations of an ionic liquid in a slit nanopore. Before we proceed, it is necessary to emphasise the following. Firstly, in the Bethe-lattice approach the dimensionality of the system does not appear explicitly but enters the model only via a coordination number $q$ (note that $q$ can be the same in different dimensions, or different for different structures in the same dimension). In simulations we consider a three-dimensional system, but restrict the ions to live inside slit-shaped ultra-narrow pores, which shall effectively reduce the coordination number as compared to the bulk. Secondly, formulating the model on a lattice, we implicitly imposed the structure which the ionic liquid attains in the ordered or disordered state. For off-lattice systems, the properties of the these phases are {\em a priori} unknown and the purpose of our Monte Carlo simulations is to understand the structure of an ionic liquid in such a strong nano-confinement.

\subsection{Simulation method}
\label{sec:sim:method} 

\begin{figure}
\includegraphics[width=.45\hsize]{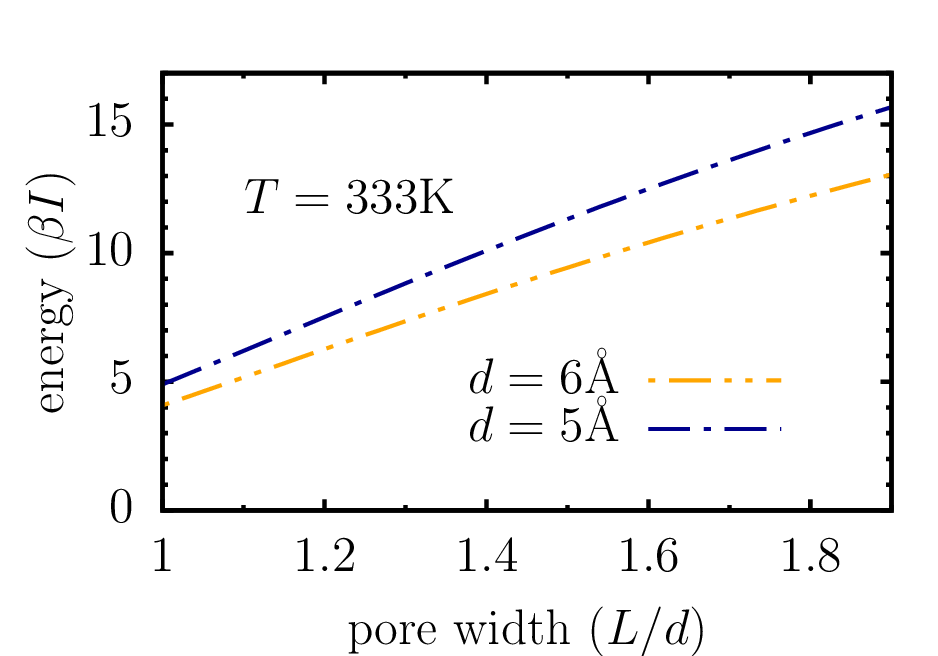}
\caption{Relation between the pore width and the coupling constant $\beta I = \beta I_{++} = \beta I_{--} = - \beta I_{+-}$ of the Hamiltonian (\ref{hamil}). Interaction potential~(\ref{eq:potential}) at the closest contact for ions located on the symmetry plane of the pore has been used to calculate $\beta I$.}
\label{fig:pot}
\end{figure}

Ionic liquid molecules were modelled as charged hard spheres and a pore was constructed from two parallel metal hard walls placed distance $L$ apart. For the ion-ion interaction potential we take
\begin{align}
\label{eq:potential}
v_{\alpha\beta}(z_1, z_2, r) = 
	\frac{4q_\alpha q_\beta }{\varepsilon_p L} 
		\sum_{n=1}^{\infty} K_0(\pi n r /L)
			\sin(\pi nz_1/L)\sin(\pi n z_2/L).
\end{align}
where $q_{\alpha}$ and $q_\beta$ are ion charges, $r$ the lateral distance between the ions, $z_1$ and $z_2$ $\in [0,L]$ are their positions perpendicular to the pore walls, and $\varepsilon_p$ the dielectric constant inside the nanopore. In what follows we take a constant, pore-width independent $\varepsilon_p = 2.42$, but we note that $\varepsilon_p$ can in principle depend on $L$ and this may have a profound effect on the system behaviour,~\cite{kondrat:ec:13} particularly on the dependence of the coupling constant of our lattice model, $\beta I$, on the slit width (\cf Figure~\ref{fig:pot}).

Interaction potential (\ref{eq:potential}) follows from the exact solution of the electrostatic problem of a point charge confined between metal walls\cite{kondrat:jpcm:11} and determines the coupling constants $I_\pm $ which thus depend on the pore width and ion diameter. Figure~\ref{fig:pot} shows $\beta I = \beta|I_\pm|$ as a function of pore width for ions located on the central symmetry plane at the closest contact. This figure suggests that in realistic systems only a first order (discontinuous) phase transition can potentially be observed. Indeed, as shown by our Bethe-lattice approach, continuous (second-order) transitions may take place only at low values of $\beta I$ (see Figure~\ref{fig:phase_diagram}), which do not seem to be typical for ionic liquids in nanopores. Continuous transitions, however, can not in general be ruled out for other systems.

Ion-pore wall potential due to the charge--image-charge interactions are (for monovalent ions)\cite{kondrat:jpcm:11}
\begin{align}
\label{eq:self}
E_\mathrm{self} (z)  = - \frac{e^2 }{\varepsilon_p L} \int_0^\infty\left[\frac{1}{2} - \frac{\sinh(k(1-z/L))\sinh(kz/L)}{\sinh(k)}\right]dk,
\end{align}
where $z$ is the position across the pore and $e$ the elementary charge. These interactions are defined as the difference between the electrostatic self energy of a point charge inside and outside of a pore. It does not depend on ion densities and the sign of the charge, and it is negative for $0 \le z \le L$ promoting ions to enter a pore. 

In simulations, similarly as in the lattice model, the resolvation energy of ions ($w_\mathrm{sim} = w^\mathrm{sim}_\pm$) controls the ion density in the pore. In the lattice model, the resolvation energy~\footnotemark[1] $w$ (see Eq.~(\ref{eq:bethe:mu})) contains the ion--pore-wall interactions due to the image-forces. This means that the two are related by $w_\mathrm{sim} = w + E_\mathrm{self} (z=L/2)$, assuming that ions position themselves on the symmetry plane of the pore. For instance, for a $0.55$nm wide pore the shift in the resolvation energy is $E_\mathrm{self} (L/2) \approx - 25k_BT$.

Potentials (\ref{eq:potential}) and (\ref{eq:self}) have been implemented in Towhee simulation package~\cite{TOWHEE, martinMCCCSTowhee:2013} and grand canonical Monte Carlo simulations have been performed using the standard translational move, Widom insertion/deletion move~\cite{widomMC:1963} and molecule-type swap move.~\cite{kondrat:pccp:11} Periodic boundary conditions were applied in the lateral ($x$ and $y$) directions. A single simulation consisted of in total $5\times 10^6 - 10^7$ steps in equilibration runs and $2 \times 10^7 - 5 \times 10^7$ in production runs. For dense systems we performed a second round of simulations starting from the saved molecular configurations obtained in the previous runs.

\subsection{Monte Carlo results}

\begin{figure}
\includegraphics[width=.45\hsize]{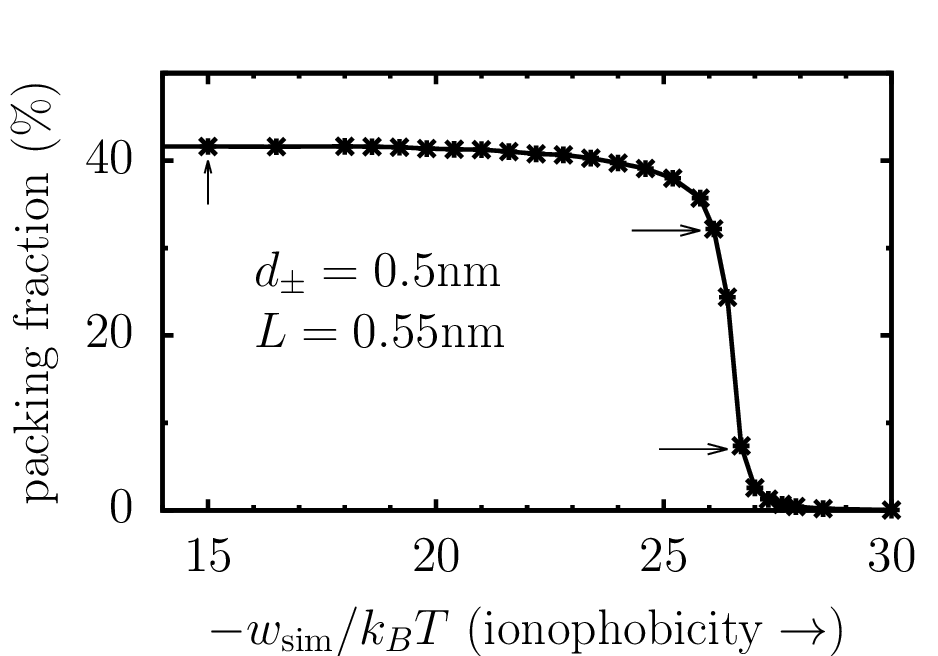}
\caption{Total ion density in a slit pore as a function of the resolvation energy $w$ (here negative of ionophobicity) obtained from Monte Carlo simulations. Ion diameter $d=0.5$nm, pore width $L=0.55$nm, and temperature 333K. The arrows point out the values of $w$ shown in Figure~\ref{fig:mcsim:struct}. The resolvation energy in the simulations is related to the resolvation energy of the lattice model by $w_\mathrm{sim}  = w + E_\mathrm{self} (L/2) \approx w - 25k_BT$. This is because $w$ contains the ion-pore interaction energy (\ref{eq:self}), due to the ion self-energy, which was not included in $w_\mathrm{sim}$.}
\label{fig:mcsim:density}
\end{figure}

It is not easy to identify the order parameter ($\delta \rho$) from our Monte Carlo simulations, and we show therefore the total ion density as a function of the resolvation energy $w_\mathrm{sim}$, the negative of which is called ionophophicity~\footnote[1]{For convenience of analytical calculations, the definition of the resolvation energy, $w_\pm$, is taken here with the sign that is opposite to the sign of the resoltation energy used in our recent works.~\cite{kondrat:nh:16, lee:16:hyster} In particular, $w_\pm^\mathrm{sim}$ of our Monte Carlo simulations is $-\delta E_\pm$ of Refs.~\onlinecite{kondrat:pccp:11, kondrat:pccp:11, kondrat:ec:13, kondrat:nh:16, lee:16:hyster}. (As noted, $w_\pm^\mathrm{sim}$ is shifted with respect to $w_\pm$ of the lattice model because of the ion self energy, Eq.~(\ref{eq:self}), which is taken into account exactly in the simulations and which depends on the ion position across the pore.) The sign of the resolvation energy $w_\pm$ in this work is however the same as in Refs.~\onlinecite{lee:prl:14,rochester:1d}, where exactly solvable Ising-like models have been adopted to study charging of cylindrical pores. The ionophobicity is defined here as the negative of $w_\pm$, while ionophilicity is naturally associated with $w_\pm$. } (see Figure~\ref{fig:mcsim:density}). Although the density exhibits a similar behaviour as predicted by the theory, the transition between the dilute and dense states (or ionophobic and ionophilic, or disordered and `ordered' or `crystalline') occurs rather smoothly, and we have not found sufficiently strong arguments to identify a phase transition. It is possible that this is due to the absence of true long-range order in two dimensional `solids' (note however that our system is only quasi two-dimensional), so that the ionic liquid transforms smoothly into the locally ordered state, but remains fluidic on a larger scale, as the pore ionophilicity increases (see also below). On the other hand, phase transitions are associated with singularities in the free energy which are not easy to capture in simulations.\cite{rotenberg_salanne:jpcl:15} 

\begin{figure}
\includegraphics[width=\hsize]{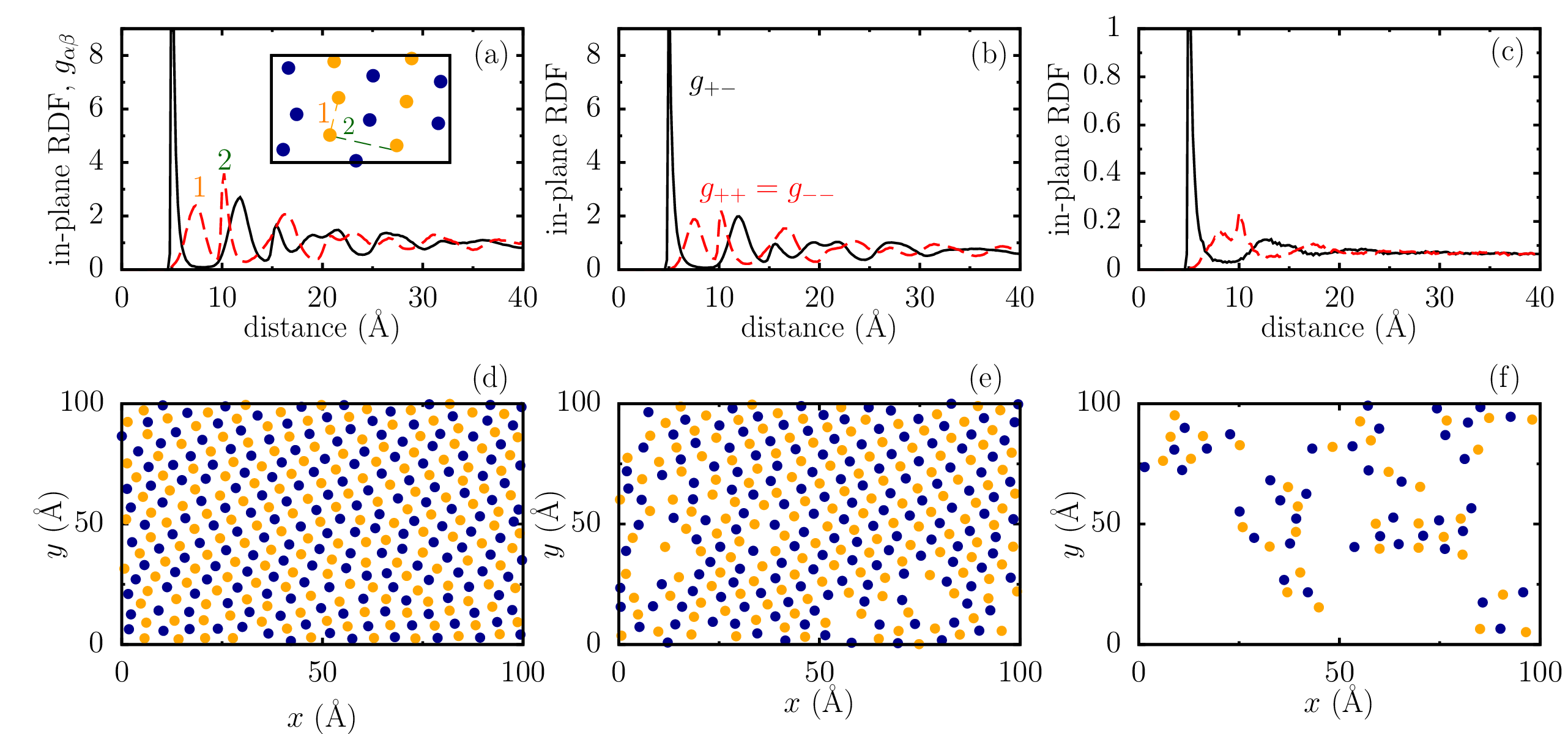}
\caption{Unnormalized in-plane radial distribution functions (RDFs) $g_{\alpha\beta}$ for (a) $w_\mathrm{sim} = -15 k_BT$, (b) $w_\mathrm{sim} = - 25.8k_BT$ and (c) $w_\mathrm{sim} = - 26.7k_BT$. These values are shown by arrows in Figure~\ref{fig:mcsim:density}. The solid and dash lines show cation-anion and cation-cation RDFs, respectively. The anion-anion RDFs coincide with the cation-cation RDFs and are not shown. The inset in (a) shows a snapshot from our Monte Carlo simulations and demonstrates the origin of two peaks in the cation-cation (and anion-anion) RDFs: The first peak denoted by (1) comes from the cations from the same `cationic snake', while the second peak denoted by (2) originates from the cations from the neighboring snakes. Temperature is $333$K, ion diameter $0.5$nm, and pore width $0.55$nm. The lower plots (d-f) show the corresponding snapshots from the Monte Carlo simulations. 
}
\label{fig:mcsim:struct}
\end{figure}

Figure~\ref{fig:mcsim:struct} shows the cation-anion radial distribution functions ($g_{+-}$) for three values of the resolvation energy $w_\mathrm{sim}$. The first peak in all cases is at $r \approx d_\pm = 5$\AA~and suggests formation of ion pairs.\cite{gebbie_israelachvili:pnas:13:ILdilute} The system exhibits a short-range order in a dense state (small $-w_\mathrm{sim}$) that extends to about $30$\AA, but its magnitude decreases with decreasing $w_\mathrm{sim}$, as one may expect (compare Figure~\ref{fig:mcsim:struct}a-b). For a dilute state, corresponding to a weak ionophobicity, the cation-cation RDF shows a behaviour typical for liquids.

The cation-cation (and anion-anion) RDFs show unusual two peaks (dash red lines in Figures~\ref{fig:mcsim:struct}a-c), instead of the standard single peak located between the subsequent peaks in $g_{+-}$. These peaks in $g_{++}$ can be related to the formation of ion `snakes' of the same sign; these snakes result, in fact, from the cation-anion chains which are shifted with respect to each other such that the cations and anions form separate snakes (see Figures~\ref{fig:mcsim:struct}d-f).  The first peak in $g_{++}$ is likely due to the in-snake neighboring cations and is located at $r \approx 7.4$\AA~$\gtrsim d$ ($d=d_\pm$ is the ion diameter). The second peak is at $r \approx 10.4$\AA~$\approx 2d$ and comes from the cations from two different cation-snakes separated by an anion snake (see the inset in Figure~\ref{fig:mcsim:struct}a). Remarkably, this two-peak behaviour survives also at low densities, although it is less pronounced (Figure~\ref{fig:mcsim:struct}c). A similar snake-like structure has been obtained within a continuous mean-field theory for molten salts confined between electrodes of different polarity.\cite{gavish_yochelis:jpcl:16}

Figures~\ref{fig:mcsim:struct}d-f show the snapshots from the Monte Carlo simulations. Interestingly, as the resolvation energy decreases and the pore becomes less occupied by ions, first \emph{clusters of voids} appear in the system which take up more and more space until they occupy most of the space and we see small mostly neutral clusters of an ionic liquid (compare Figures~\ref{fig:mcsim:struct}e and f). A similar clusterisation has also been observed for ionic liquids in the bulk~\cite{bernardes:jpcb:11:ILClusters, chen:pccp:14:ILclusters}. It is difficult to estimate the life-time of these clusters from our Monte Carlo simulations, but they seem to be relatively stable and appear in all snapshots we have looked at. Such a clusterisation may have important implications for charging kinetics and deserves a separate study.

\section{Conclusions and discussion}
\label{sec:concl}

We have studied the phase behaviour and structure of ionic liquids confined in non-polarised narrow slit pores with conducting walls. In such a confinement, the interactions between the ions are effectively screened out, and we have used this fact to formulate and solve a lattice model of such a \emph{superionic} liquid, taking into account the nearest neighbour interactions only, and resorting to the Bethe-lattice approach for bipartite lattices with the number of nearest neighbours $q=3$ and $q=4$. This approach has been extensively used in different contexts and for various lattice models, and has shown to reproduce well the phase behavior, including the order of phase transitions, and to give a reasonable estimate for the location of phase transitions. We supplemented these analytical results by off-lattice Monte Carlo simulations of an ionic liquid in slit narrow pores.

Within the Bethe-lattice approach, we calculated the complete phase diagram of a superionic liquid in the $(\beta I, \beta w)$ plane (Figure~\ref{fig:phase_diagram}), where $\beta $ is the reciprocal temperature, $I $ is the (pore-width dependent) strength of the ion-ion interactions and $w $ the ion's resolvation energy determining the affinity of ions towards pores. The phase diagram consists of a disordered phase, in which one has a homogeneous mixture of ions of two types and voids, and an ordered phase, in which ions of one type occupy predominantly one of the sub-lattices. These two phases are separated by a demarcation line which approaches a finite asymptotic value $(\beta I)_\mathrm{threshold}$ as the resolvation energy $\beta w \to +\infty$ (implying strongly ionophilic pores). No phase transition takes place for $\beta I < (\beta I)_\mathrm{threshold}$ and the system is in the disordered state. For $\beta I $ between  $(\beta I)_\mathrm{threshold}$ and  $(\beta|I|)_{tc}$, we observe a line of critical points (second order transitions) terminating at the tricritical point $(\beta I)_{tc}$. It is possible, however, that fluctuations (and long-range interactions) can shift this transition upwards and make it first order.~\cite{brazovskii:jetp:75} For a second order transition, the total ion density and the order parameter vary continuously across the transition; both quantities show a cusp and the quantity analogous to compressibility exhibit a finite jump at the transition (Figures~\ref{fig:bethe:op} and \ref{fig:compres}a, respectively). The first order transition is predicted for $\beta I$ above $(\beta|I|)_{tc}$. In this case, there is a finite jump in the total ion density, order parameter and compressibility (Figures~\ref{fig:bethe:op} and \ref{fig:compres}b), and the compressibility shows additionally a spike at the transition.

An important observation is that by increasing the coordination number ($q$), the line of phase transitions between the disordered and ordered phases shifts down left, and the value of $(\beta I)_\mathrm{threshold} = \ln \sqrt{q/(q-2)}$, below which no transition occurs, decreases with increasing $q$. This implies that systems with larger coordination numbers enter into the disordered state at lower values of $\beta I$ and $\beta w$, and thus at higher temperatures (at fixed $I$ and $w$). This is understandable because higher energies are required to break cation-anion `bonds' in the ordered state, as compared to a system with fewer such bonds, whose number increases with increasing the coordination number $q$.

We have realised that in realistic ionic liquids in slit nanopores, the interaction strength ($\beta I$) is typically of the order of a few $k_B T$ or more (Figure~\ref{fig:pot}). Since the tricritical points are at rather low values of $\beta I$ ($(\beta I)_{tc} \approx 1.8$ for $q=3$ and $(\beta I)_{tc} \approx 1.09$ for $q=4$), it seems unlikely to observe continuous transitions between the ordered and disordered states in confined ionic liquids. We may however expect to see the first order transitions, which are characterised by the discontinuity of the density at the transition, and can potentially be seen experimentally by changing temperature or pore width. Interestingly, recent experimental~\cite{griffin:natmat:15, force:jacs:NMRDyn} and theoretical~\cite{rochester:1d} studies suggest that ionophobicity of pores can be effectively controlled by using solvent. This means that such phase transitions can potentially be observed as a function of the solvent concentration.

Surprisingly at the first glance, our theory predicts that the transition into the disordered phase happens at negative values of the resolvation energy~\footnotemark[1] $w$ (see Eq.~(\ref{eq:bethe:mu})), which also means that a transformation from ionophilic to ionophobic occurs at $w<0$ (unlike in simulations where it is at $w\approx 0$, see Figure~\ref{fig:mcsim:density}). This result can be understood as follows. For a strongly ordered phase, sub-latticing implies that each ion faces only ions of the opposite sign (see Figure~\ref{fig:honeycomb}), and we can easily estimate the chemical potential $\mu_\textrm{ordered} \approx - Iq/2 - w$. For the disordered phase and low values of $w$ we have $\mu_\mathrm{disordered} \approx 0$, and we thus find that the (first order) transition occurs at $w_\mathrm{transition} \approx -Iq/2$. This estimate agrees remarkably well with the exact result (Figure~\ref{fig:phase_diagram}). However, our simulations show that ions have both cations and anions as their neighbours (see Figure~\ref{fig:mcsim:struct}d-e), and this shall decrease the contribution from the ion-ion interactions (the first term in $\mu_\mathrm{ordered}$) and thus increase $w_\mathrm{transition}$. Additionally, next to nearest (and higher order) neighbour interactions, neglected in our lattice model, will reduce the value of $w_\mathrm{transition}$ further, and may potentially bring it closer to zero (as seen in simulations). It would thus be very interesting to study the effects of such long-ranged interactions on the location, order and existence of the phase transitions predicted here by the Bethe-lattice approach.

Our Monte Carlo simulations support the Bethe-lattice results in that they (i) demonstrate an abrupt drop in the ion density with an increase of ionophobicity (compare Figures~\ref{fig:bethe:op}b and \ref{fig:mcsim:density}); and (ii) for a dense state they show the existence of two `sub-lattices' of anions and cations and in-plane crystal-like (albeit short range) ordering of ions (Figure~\ref{fig:mcsim:struct}). However, due to the finite size of a simulation box and limited computational resources, it has not been possible to reproduce the full `phase diagram' or even find sufficiently strong arguments to identify a true phase transition. Further work is therefore required to verify the predictions of our lattice model. 

Interestingly, however, the simulations reveal the formation of ordered ionic liquid clusters, separated by voids, in the region on the phase diagram where the lattice model predicts the ordered phase. Such a clusterization might be a sign of the onset of a phase separation between the dense (ordered) and dilute (homogeneous or disordered) phases, and may have important consequences for charging dynamics. Finally, our Monte Carlo simulations suggest formation of ionic `snakes' which lead to  unusual two peaks located in close proximity of each other in the cation-cation and anion-anion radial distribution functions (Figures~\ref{fig:mcsim:struct}d-f). It would be very interesting to verify these findings experimentally, \textit{e.g.}~by neutron diffraction on ions using isotopic substitution (provided of course that a large contrast with electrodes can be achieved), and thus to shed new lights on the structure and phase behaviour of ionic liquids in such strong nanoconfinements.

\appendix

\section{Mapping to the classical spin S = 1 model}
\label{app:map2BEG}

Since  the hard-core interaction excludes the state $(n_i,m_i) = (1,1)$ for the same site, there is a  well-known connection between the present model and the three-state lattice gas model \cite{3state} or, equivalently, a special Blume-Emery-Griffiths (BEG) spin $S = 1$ model \cite{BEG}. The mapping to the BEG model is accomplished as follows. We assign, in a usual fashion,  to each site a three-state variable $S_i$, such that
\begin{equation}
S_i =
\begin{cases}
+1, & ~~\textrm{site $i$ is occupied by `$+$' particle,} \cr
-1, & ~~\textrm{site $i$ is occupied by `$-$' particle,} \cr
~~0,  & ~~\textrm{site $i$ is  empty.}\cr
\end{cases}
\label{spin}
\end{equation}
Then, the occupation numbers $n_i$ and $m_i$ may be straightforwardly 
rewritten  in terms of $S_i$ as
\begin{eqnarray}
n_i=(S_i+S_i^2)/2\qquad m_i=(-S_i+S_i^2)/2
\end{eqnarray}
so that the Hamiltonian becomes
\begin{eqnarray}\label{BEG}
{\mathcal H}=-J \sum_{\langle ij\rangle}S_i S_j 
	- K \sum_{\langle ij\rangle}S^2_i S^2_j
	- C \sum_{\langle ij\rangle}(S_iS^2_j+S_j S^2_j)
	-h \sum_{i} S_i + \Delta\sum_{i}S^2_i,
\end{eqnarray}
where
\begin{eqnarray}
J&=&-\frac{\left(I_{++}+I_{--}+2I_{+-}\right)}{4},~K=-\frac{\left(I_{++}+I_{--}-2 I_{+-}\right)}{4},~C=-\frac{\left(I_{++}-I_{--}\right)}{4},\nonumber\\
h&=&\frac{\mu_+^{(0)} -\mu_-^{(0)}}{2},~\Delta=-\frac{\mu_+^{(0)} +\mu_-^{(0)}}{2}.
\end{eqnarray}
As one can readily notice, $J<0$, since all interactions strengths $I$ are positive in our case.
Note that such model with $J<0$ was already applied  to describe crystallization and the order-disorder transition  in a binary alloy within a mean-field approximation\cite{Saito81}.

In case of equal interaction strengths between the like species, \ie, when $I_{++}=I_{--}$,  the constant $C$ in (\ref{BEG}) becomes equal to zero, $C = 0$, so that the model reduces to a well-studied  version of the original BEG model. It was treated by a wide variety of approaches (see, e.g., Ref.~\onlinecite{beg_ref} and references therein) and has a rich phase diagram including paramagnetic, ferromagnetic, quadrupolar phases, and multicritical points, depending on the ratio of $K/J$. Note that the ferromagnetic case $J>0$ was mainly studied for bipartite lattices since in the absence of an external field  ($h=0$) one can map  the antiferromagnetic case $J<0$  onto the  ferromagnetic case  by merely redefining the spin directions on one of the sublattices. Note that in ionic liquids ions are single charged in the vast majority of cases. Therefore, the choice  $I_{++}=I_{--}$ is absolutely ``natural'', given that the ionic valencies are equal.

Physically, for the systems with Coulomb interactions, when $I_{++}=I_{--}$, assuming also that the ions of both types have approximately the same size, one should expect that $I_{++}=I_{--}=I_{+-}=I$. This leads to $K=C=0$, and thereby  to the so-called spin-1 Blume-Capel (BC) model in a magnetic field. This model was originally invented to describe magnetic systems \cite{Blume66,Capel66,*Capel67,*Capel67b} and has been subsequently applied to a large variety of physical problems (see, \textit{e.g.}, Ref.~\onlinecite{Lawrie88} and references therein). The original mean-field treatments \cite{Blume66,Capel66,*Capel67,*Capel67b} were continued \cite{Wang78,Plascak93} and completed by various analytical and numerical studies of two-dimensional and tree-dimensional BC models \cite{Oitmaa70,*Oitmaa71,*Oitmaa72, Brankov72,Saul74,Siqueira86,Kaneyoshi86, *Kaneyoshi86b, *Kaneyoshi90, Costabile12, *Costabile14, Micnas79,Bonfim85,Burkhardt76, *Burkhardt77,nonperturbative,*nonperturbative2,Arora73,Jain80,Kimel86,Wang91,Kimel92,Care93,Desarno97,PenaLara98,Pawlowski06,Silva06,Takana,Beale,Ekiz01}. We note that already a mean-field analysis  presented in the seminal works  \cite{Blume66,Capel66,*Capel67,*Capel67b} showed that the BC spin-1 model exhibits a second-order phase transition line separating a disordered phase from an ordered one, and changing at a tricritical point into the line of the first-order phase transitions for sufficiently large values of $\Delta$. The phase diagram, location of the tricritical point, as well as values of the critical exponents were quantitatively analysed within high- \cite{Oitmaa70,*Oitmaa71,*Oitmaa72,Brankov72,Saul74},  and low-temperature series expansion methods \cite{Saul74}, different effective theories \cite{Siqueira86,Kaneyoshi86, *Kaneyoshi86b, *Kaneyoshi90,Costabile12, *Costabile14}, variational approximations \cite{Micnas79}, mean-field renormalization group (RG) \cite{Bonfim85}, Kadanoff’s lower-bound RG transformations \cite{Burkhardt76, *Burkhardt77}, nonperturbative RG schemes \cite{nonperturbative, *nonperturbative2},  various Monte-Carlo methods \cite{Arora73,Jain80,Kimel86,Wang91,Kimel92,Care93,Desarno97,PenaLara98,Pawlowski06,Silva06} , constant-coupling approximation \cite{Takana}, transfer matrix finite-size scaling\cite{Beale}, lowest approximation of cluster variation method \cite{Ekiz01}, and pair approximations for the free energy \cite{PenaLara98}.

We note finally that the BC model has also been studied using the Bethe lattice approximation. The first exact results for the BC model on the Bethe lattice with a general coordination number $q$  were obtained within the BEG model \cite{Olivera85,Chakraborty84,Chakraborty86,Ananikian91}. Analysing fixed points of exact recurrent equations, the  specific features of the transition between the paramagnetic and the nferromagnetic phases, as well as the existence of the tricritical point were investigated.  As was noted in Ref.~\onlinecite{Osorio89}, these studies did not  take  into account a bipartite nature of the Bethe lattice. Recursion relations for the bipartite lattices in addition  to
 fixed points have also  cycles,  associated with different thermodynamical phases. Correlation functions for the  BEG model on Bethe lattice were obtained in Refs.~\onlinecite{Hu98,Izmailian98}. Numerical analysis of fixed points and cycles solutions of exact recursion relations were performed in Refs.~\onlinecite{Akheyan96,Erdinc2006}. For the case of Blume-Capel model such investigations also were performed \cite{Ekiz2004}.

\section{Bethe lattice solution for coordination number $q=3$ (details)}
\label{sec:q3:details}

Using variables $u$ and $v$, Eq.~(\ref{eq:u_v}), the total ion density and the order parameter on one of the sublattices, Eq.~(\ref{eq:op:def}), can be put in the form:
\begin{align}
\label{eq:uv}
 \bar\rho^A =\frac{2uz(u^2+3v^2)}{1+2uz(u^2+3v^2)},\quad
 \delta\rho^A =\frac{2vz(3u^2+v^2)}{1+2uz(u^2+3v^2)}
\end{align}
From (\ref{full}) and (\ref{function}), we find that $u$ and $v$  obey the following system of two coupled nonlinear equations:
\begin{eqnarray}\label{uv}
u&=&\cosh
   ({\beta I})+\frac{({F}+1)^2 (1-\cosh ({\beta I}))}{16 u^2 z^2 \sinh ^2({\beta I}) \left({F}-2 u^2 z\right)+2 z ({F} \cosh
   ({\beta I})+1)^2+({F}+1)^2};\nonumber\\
v&=&\frac{16 u v z^2 \sinh ^2({\beta I}) ({F} \cosh ({\beta I})+1)}{16 u^2 z^2 \sinh ^2({\beta I}) \left({F}-2 u^2 z\right)+2 z ({F} \cosh
   ({\beta I})+1)^2+({F}+1)^2},
\end{eqnarray}
where
\begin{eqnarray}
F=2 z (u^2+v^2) \,.
\end{eqnarray}
This system of equation has two sets of solutions: (a)~$v=0$, corresponding to the homogeneous case with $x_A=y_A=x_B=y_B$ and (b)~$v \neq 0$, in which case the symmetry between the two sub lattices is broken.

In the case (a), the variable $u$  obeys the following non-linear equation
\begin{equation}
f_1(u,z,I) f_2(u,z,I)=0,
\end{equation}
with
 \begin{equation}
f_1(u,z,I) = u - 1 + 2 u^3 z \left (1 - \cosh(\beta I)\right) 
\end{equation}
Equation $f_1=0$, which is cubic in $u$, has a single real solution, which is depicted  in Figures~\ref{Fig1} and~\ref{fig:f1f4} by a solid black curve. This solution can also be obtained from the first equation in (\ref{A}) by setting $x_B=y_B=x_A=y_A$. It describes a disordered phase, where we have the same densities of '+' particles and '--' particles on each of the sublattices, and in consequence - in the whole system. In this case one evidently has that $\delta\rho^A=\delta\rho^B=0$ and $\bar\rho^A=\bar\rho^B=\bar\rho$.

Next, the function $f_2(u,z,I)$ is given explicitly by
 \begin{equation}
f_2(u,z,I)=1 + 2 (u-1) u z + 2 z \cosh(\beta I) (1 + u + 2 u^2 z \cosh(\beta I))
\end{equation}
 and describes the situation when $x_A=y_A\not=x_B=y_B$. Equation $f_2(u,z,I)=0$ has no real solutions, which means that thermodynamical phases with $x_A=y_A\not=x_B=y_B$ do not exist.

In the case (b), \ie, for  $v\not =0$, we can express $v$ through the variable $u$ from the second equation in (\ref{uv}), to get the following closed-form non-linear equation
\begin{equation}
f_3(u,z,I)f_4(u,z,I)f_5(u,z,I)=0
\end{equation}
 where
\begin{equation}\label{f3}
f_3(u,z,I)=1 - \cosh(\beta I) + 4 u z \sinh(\beta I) ( u-\cosh(\beta I) )
\end{equation}
describes the situation with the broken symmetry, that is,  $x_A=x_B\not=y_A=y_B$ with
\begin{equation}\label{f3v}
1+F + 4 u z \sinh(\beta I)=0
\end{equation}
Real solutions for $x_A$, $y_A$ are obtained in the case $I<0$, corresponding to ferromagnetic ($J>0$) BC model.  Solutions of the latter equations are shown by red line in Figure~\ref{Fig1}. This solution can be also obtained from  (\ref{A}) by setting $x_B=x_A$ and $y_B=y_A$.
Next, we have
\begin{equation}\label{f4}
f_4(u,z,I)=1 - \cosh(\beta I) - 4 u z \sinh(\beta I) ( u-\cosh(\beta I) ) \,,
\end{equation}
which describes the case $x_A=y_B\not=x_B=y_A$ with
\begin{equation}\label{f4v}
1+F - 4 u z \sinh(\beta I)=0.
\end{equation}
 Real solutions for $x_A$ and $y_A$ are obtained in the case $I>0$, which now
 corresponds to antiferromagnetic ($J<0$) BC model.
Note that (\ref{f3}) and (\ref{f3v}) transform into (\ref{f4}) and (\ref{f4v}) (and {\em vice versa}) upon the change of the sign, \ie,  $I\to -I$.   
It was expected for bipartite lattices that for $h=0$ results for the 
ferromagnetic model correspond to the results of the antiferromagnetic model upon the reversal of the sign of the interactions, $J\to -J$ \cite{Olivera85,Akheyan96}.  However, explicit  solution for the 
antiferromagnetic model can be extracted using the equations for the sublattices, as we have shown. 
Real solutions for $x_A$ and $y_A$ are depicted by the red line Figure~\ref{Fig1}. These real solutions describe the ordered phase, where we have the same densities of '+' particles on the sublattice A, and of  '--' particles on the  sublattice B (and {\em vice versa}). 
 Densities  $\rho_+$ and $\rho_-$  in the whole system are the same.
Therefore, in this case we have $\delta\rho^A=-\delta\rho^B$ and $\bar\rho^A=\bar\rho^B=\bar\rho$.

Finally, the function  $f_5(u,z,I)$ describes the case $x_A\not=y_A\not=x_B\not=y_B$ and is defined explicitly by 
\begin{eqnarray}
\!\!f_5(u,z,I)&{=}&-16 u^4 z^4 \sinh ^4(2 {\beta I})-128 u^2 z^3 \sinh ^4({\beta I}) \cosh ({\beta I}) \left(\cosh ^2({\beta I})+u^2\right)-\nonumber\\&&16 u^2 z^2 \sinh ^2({\beta I}) \left((\cosh
   ({\beta I}){-}1)^2 \left(\cosh ^2({\beta I}){-}2 u \cosh ({\beta I}){-}u^2\right){+}4 \sinh ^2({\beta I})\right){+}\nonumber\\&& 16 u z \sinh ^2({\beta I}) (\cosh ({\beta I})-1)^2-(\cosh
   ({\beta I})-1)^4
\end{eqnarray}
Equation $f_5 = 0$ does not have any real solution for $x$ and $y$.

\section{Bethe lattice solution for coordination number $q=4$}
\label{sec:q4}

\begin{figure}
\includegraphics[width=.42\hsize]{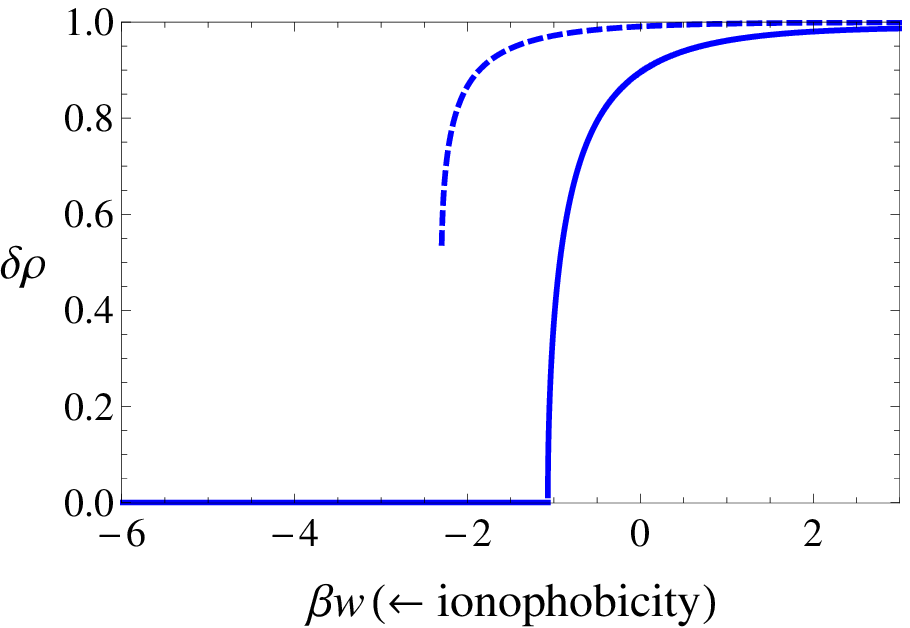} 
\includegraphics[width=.42\hsize]{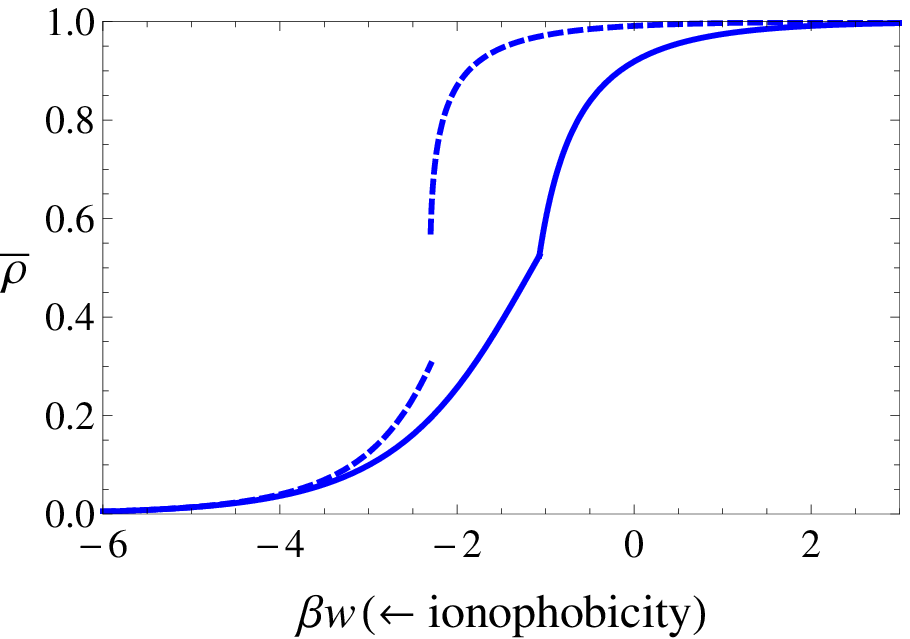}

\hspace{1cm}\mbox{(a)}\hspace{6cm}\mbox{(b)}
\caption{
(a)~The order parameter, $\delta\rho$, and (b) the total ion density, $\bar\rho$, as a function of resolvation energy $\beta w$. The results are obtained by the Bethe-lattice approach for coordination number $q=4$.  Solid lines show $\delta \rho$ and $\bar \rho$ for a second order transition ($\beta I=0.7$) and the dash lines for a first order transition ($\beta I=1.2$).
}
\label{Fig6}
\end{figure}

For coordination number $q=4$, the disordered phase is described by the equations
\begin{eqnarray}
v&=&0\\
f_1(u,z,I)&=&1+2 u^3 z \cosh (\beta I)-u \left(2 u^3 z+1\right)=0,
\end{eqnarray}
while for the ordered phase we have
\begin{eqnarray}
2 u z \left(u^2+3 v^2\right) (\sinh (\beta I)-3 u)+u \left(16 u^2 z \sinh (\beta I)-3\right)&=&0\\
f_4(u,z,I)=(\cosh (\beta I)-u) \left(3-16 u^3 z \sinh (\beta I)\right)-(u-1) (\sinh (\beta I)-3)&=&0,
\end{eqnarray}
From these equations we obtain for the second order phase transitions:
\begin{equation}
z_c =	\frac{ (-3 \sinh ((\beta I)_c)+\cosh ((\beta I)_c)-1)^3}{54  \sinh ((\beta I)_c)^3(\cosh ((\beta I)_c)-3 \sinh ((\beta I)_c))},
\end{equation}
which are depicted by a solid line in Figure~\ref{fig:phase_diagram}. The order parameter $\delta\rho$  as well as density of all '+' and '--' particles $\bar\rho$  are shown in Figure~\ref{Fig6}.

It is important to note that we observe essentially the same qualitative behavior of the pertinent parameters as in the case of the Bethe lattice with the coordination number $q  = 3$. This suggest that our conclusions are likely generic and are expected to be valid for an arbitrary coordination number of the embedding lattice.


%

\end{document}